\documentclass[sigconf, nonacm]{acmart}
\usepackage{hyperref}
\usepackage{enumitem}
\usepackage{graphicx}
\usepackage{adjustbox}
\usepackage{array}
\usepackage{booktabs}
\usepackage{amsmath}
\usepackage{geometry}
\usepackage{graphicx}
\usepackage[utf8]{inputenc}
\usepackage{xcolor}
\usepackage{tcolorbox}
\usepackage{lipsum}
\usepackage[T1]{fontenc}
\usepackage{float}

\usepackage{listings}

\usepackage{caption}
\usepackage{multirow}
\usepackage{makecell}
\usepackage{longtable}

\usepackage{colortbl}

\usepackage{colortbl}
\usepackage{booktabs}
\usepackage{algorithm}
\usepackage{algpseudocode}
\usepackage{tabularx}
\usepackage{placeins}
\definecolor{lightgray}{gray}{0.95}
 \usepackage{algorithm}
\usepackage{algpseudocode}
\usepackage{fancyhdr}
\usepackage{pifont}
\usepackage{listings}

\definecolor{lightgray}{gray}{0.97}
\definecolor{commentgreen}{rgb}{0,0.5,0}
\definecolor{tagblue}{rgb}{0.1,0.1,0.8}
\definecolor{attrpurple}{rgb}{0.58,0,0.82}
\definecolor{stringred}{rgb}{0.7,0,0}

\fancypagestyle{nofooter}{
  \fancyhf{} 

}

\newcommand{\method}{\emph{AccessGuru}} 
\makeatletter
\renewcommand\subsubsection{\@startsection{subsubsection}{3}{\z@}%
  {-18\p@ \@plus -4\p@ \@minus -4\p@}%
  {-\p@}%
  {\normalfont\normalsize\itshape}%
}
\makeatother

\AtBeginDocument{%
  }

\copyrightyear{2025}
\acmYear{2025}
\setcopyright{none} 
\acmConference[ASSETS '25]{Preprint accepted at ASSETS '25}{October 26--29, 2025}{Denver, CO, USA}
\acmBooktitle{Preprint accepted at The 27th International ACM SIGACCESS Conference on Computers and Accessibility (ASSETS '25), October 26--29, 2025, Denver, CO, USA}
 \acmDOI{}  
 \acmISBN{}  

\begin{document}

\title{AccessGuru: Leveraging LLMs to Detect and Correct Web Accessibility Violations in HTML Code}

\author{Nadeen Fathallah}
\orcid{0000-0001-7921-034X}
\email{nadeen.fathallah@ki.uni-stuttgart.de}
\affiliation{
  \institution{University of Stuttgart, Artificial Intelligence}
  \city{Stuttgart}
  \country{Germany}
}

\author{Daniel Hernández}
\orcid{0000-0002-7896-0875}
\email{daniel.hernandez@ki.uni-stuttgart.de}
\affiliation{
  \institution{Artificial Intelligence, University of Stuttgart}
  \city{Stuttgart, Baden-Württemberg}
  \country{Germany}
}

\author{Steffen Staab}
\orcid{0000-0002-0780-4154}
\email{steffen.staab@uni-stuttgart.de}
\affiliation{
  \institution{University of Stuttgart}
  \city{Stuttgart}
  \country{Germany}
}
\affiliation{
  \institution{University of Southampton}
  \city{Southampton}
  \country{United Kingdom}
}

\newcommand\freefootnote[1]{%
  \let\thefootnote\relax%
  \footnotetext{#1}%
  \let\thefootnote\svthefootnote%
}

\begin{abstract}
The vast majority of Web pages fail to comply with established Web accessibility guidelines, excluding a range of users with diverse abilities from interacting with their content. Making Web pages accessible to all users requires dedicated expertise and additional manual efforts from Web page providers. To lower their efforts and, thus, promote inclusiveness, we aim to automatically detect and correct Web accessibility violations in HTML code. While previous work has made progress in detecting certain types of accessibility violations, the problem of automatically detecting and correcting accessibility violations remains an open challenge that we address.

We introduce a novel taxonomy classifying Web accessibility violations into three key categories— Syntactic, Semantic, and Layout. This taxonomy provides a structured foundation for developing our detection and correction method and selecting and redefining evaluation metrics. We propose our novel method, \method, which combines existing accessibility testing tools and Large Language Models (LLMs) to detect accessibility violations of Web accessibility guidelines and taxonomy-driven prompting strategies of LLMs to correct all three accessibility violation categories.

To evaluate these capabilities, we have developed a novel benchmark encompassing Web accessibility violations from real-world Web pages. Our benchmark quantifies syntactic and layout compliance and judges semantic accuracy through a comparative analysis against human expert corrections. Evaluation against our benchmark demonstrates that our method achieves up to 84\% average violation score decrease on our benchmark dataset, significantly outperforming existing methods, which achieve at most 50\% average violation score decrease.
\freefootnote{%
This document is the preprint of the paper with DOI \href{https://doi.org/10.1145/3663547.3746360}{10.1145/3663547.3746360}, accepted at The 27th International ACM SIGACCESS Conference on Computers and Accessibility (ASSETS '25), October 26--29, 2025, Denver, CO, USA.
}
\end{abstract}



\maketitle

\section{Introduction} 
Web accessibility is the ability of Websites to be accessed by all people, including people with visual, auditory, motor, and cognitive impairments such that they can perceive, understand, navigate, interact with, and contribute to the Web effectively \cite{DBLP:conf/w4a/YesiladaBVH12, DBLP:conf/cscw/WuWFS17, rutter2007Web}. Organizations like Web Accessibility in Mind (WebAIM) and the World Wide Web Consortium (W3C) have established guidelines, such as the Web Content Accessibility Guidelines (WCAG) \cite{caldwell2008Web}, which advise developers to create accessible Web content. However, an overwhelming majority of Web pages do not comply with Web accessibility guidelines; for instance, a 2025 study by WebAIM revealed that 94.8\% of the top one million most visited Web pages fail to meet accessibility guidelines \cite{Webaim2025}. This widespread non-compliance is likely due to a lack of expertise in creating accessible content in the first place and the costly and labor-intensive effort of correcting accessibility issues in the HTML code \cite{DBLP:conf/icse/Alshayban0M20, DBLP:journals/tosem/BiXLGZF22}.
 
Users with impairments rely on various assistive technologies, such as screen readers for visually impaired users, screen magnification tools for individuals with low vision, closed captioning for those with hearing impairments, and voice recognition software for users with motor disabilities. These technologies function effectively when accessibility-related information is embedded in HTML. Based on the distinct types of HTML constructs and information required by assistive technologies—as specified by the WCAG—we introduce a taxonomy of accessibility violations that classifies them into three dimensions: Syntactic, Semantic, and Layout. Syntactic accessibility violations involve missing or malformed accessibility-enhancing HTML elements and attributes (e.g., missing alt text); Semantic accessibility violations concern whether the provided accessibility-enhancing HTML elements and attributes are meaningful (e.g., alt text that fails to describe the image content); and Layout accessibility violations refer to visual or structural barriers that impede interaction (e.g., insufficient color contrast). A detailed explanation of this taxonomy, its construction, and representative examples can be found in Section~\ref{sec:taxonomy}.

Research has led to the creation of efficient and accurate tools to automatically detect syntactic and layout accessibility violations in HTML, like WAVE, Axe, Google Lighthouse, and AChecker. Current tools fail to identify semantic accessibility violations such as Violation 5 (line 25) in Listing \ref{lst:accessibility_not_fixed}. This highlights a significant gap in current accessibility detection methods. 

Automating the correction of syntactic, layout, and semantic accessibility violations remains an open challenge. Web accessibility guidelines provide general recommendations rather than detailed solutions. For example, they recommend making interactive elements keyboard-operable for users without mouse access, but don't specify how to manage keyboard focus for dynamically loaded content.

This paper addresses the challenge of improving Web accessibility by introducing a method for automatically detecting and correcting HTML accessibility violations. Our taxonomy provides a systematic understanding of Web accessibility violations and informs our development of detection and correction strategies and the choice of evaluation metrics. \method \space relies on an automatic Web accessibility evaluation tool to detect syntactic and layout accessibility violations, and LLMs to detect semantic accessibility violations. \method \space transforms Web pages with accessibility violations into guideline-compliant versions by leveraging a pre-trained LLM to generate corrections that minimize a violation score, reflecting their impact on user interaction. In this paper, we make the following contributions: 
\begin{itemize} 
\item We introduce a novel taxonomy that classifies Web accessibility violations into three key categories: Syntactic, Semantic, and Layout. This taxonomy provides a structured foundation for understanding accessibility violations, guiding detection and correction strategies, and informing evaluation metrics.

\item  We present a dataset of 3,500 real-world Web accessibility violations spanning over 112 distinct types across syntactic, semantic, and layout categories, offering a diverse and representative basis for training and evaluating correction methods. To our knowledge, this is the first publicly available dataset of this scale that includes all three types of accessibility violations, sourced entirely from real-world Web data. 

\item We develop a novel pipeline that leverages accessibility testing tools and LLMs to automatically detect and correct Web accessibility violations. Building on recent ideas of using LLMs for coding \cite{DBLP:conf/icse/NamMHVM24, DBLP:conf/nips/LiuXW023, DBLP:conf/chi/Vaithilingam0G22, DBLP:conf/w4a/AljedaaniHAEF24, DBLP:journals/corr/abs-2107-03374, DBLP:journals/corr/abs-2108-07732, DBLP:conf/naacl/AhmadCRC21}.

\item We evaluate the correction effectiveness of \method \space against three baseline methods~\cite{DBLP:conf/petra/OthmanDJ23, DBLP:journals/corr/abs-2401-16450, DBLP:conf/ccnc/DelnevoAM24} using a subset of our dataset sampled to reflect real-world violation distributions as indicated by~\cite{Webaim2025}, and an additional dataset from~\cite{DBLP:journals/corr/abs-2401-16450}. Our method achieves up to 84\% average violation score decrease, outperforming baselines capped at 50\%. We compare the corrections generated by \method \space to those generated by human developers. To this end, we conducted a human developer correction study on 55 Web accessibility violations from our dataset, achieving an average similarity of 77\%.

\end{itemize}
To ensure reproducibility and transparency, all source codes, prompts, datasets, and results are available\footnote{\url{https://github.com/NadeenAhmad/AccessGuruLLM}}. The dataset is also archived with a DOI \footnote{\url{https://doi.org/10.18419/DARUS-5177}}. The structure of the paper is as follows: Section 2 provides an overview of the existing literature and background information, followed by our proposed taxonomy in Section 3. Section 4 details our approach, Section 5 describes the evaluation process, and the results from our experiments are presented in Section 6. Section 7 discusses the limitations of our method, and Section 8 concludes the paper with directions for future research.

\section{Related Work}

\subsection{WCAG Guidelines}
The WCAG \cite{kirkpatrick2023wcag,caldwell2008Web} are a widely adopted set of guidelines aimed at making Web content accessible to individuals with diverse abilities, including visual, auditory, motor, and cognitive impairments. WCAG are organized around four fundamental principles, referred to by the acronym POUR: (i) Perceivable, ensuring information and interface elements are presented in ways users can perceive, like providing text alternatives for images; (ii) Operable, requiring Website functionality to be accessible through different input methods, such as keyboard navigation or voice commands; (iii) Understandable, ensuring content is clear and easy to interact with; and (iv) Robust, which demands that content is compatible with current and future technologies.
Each of these principles is broken down into specific, testable success criteria, grouped into three conformance levels: A (minimum), AA (mid-range), and AAA (highest). 

\subsection{Web Accessibility Violation Detection}

Early efforts in accessibility violation detection relied on manual testing and expert assessment, which were time-consuming and lacked scalability. This motivated the development of efficient tools like WAVE, Axe, Google Lighthouse, AChecker, and Tenon, which translate WCAG accessibility guidelines into rule-based checks and apply them to individual HTML elements to detect syntax and layout accessibility violations \cite{DBLP:series/hci/YesiladaH19}. These reports provide developers with information on each violation, including its type, description, and impact. In our work, we utilize the Axe-Playwright tool \cite{PlaywrightAPI}, an accessibility detection engine. However, such tools fail to detect semantic accessibility violations \cite{lopez2024turning}. For instance, Axe-Playwright can confirm the presence of alt text but cannot evaluate whether the description effectively conveys the image's content. 

\subsection{Web Accessibility Violation Correction}
\emph{Rule-based Corrections of Accessibility Violations. }  
Early Web accessibility correction methods defined rules to automate corrections~\cite{DBLP:conf/hci/FeratiS16, DBLP:conf/wse/CesaranoFT07, DBLP:conf/icisdm/AlmasoudM19, tuan2012checking}. For example, \cite{DBLP:conf/hci/FeratiS16} defines a fixed rule that inserts a skip link before a navigation bar to jump to the \texttt{<main>} tag. However, such rules assume consistent page structure and fail to generalize across diverse Web layouts.

\emph{Computer Vision Correcting Accessibility Violations. } 
Computer vision techniques have been used to correct Web accessibility violations, particularly for generating alt text. Facebook's automatic alt text feature applied object detection to describe images for visually impaired users \cite{DBLP:conf/cscw/WuWFS17}, leveraging neural image captioning models such as \cite{DBLP:conf/cvpr/VinyalsTBE15}. Computer vision has been applied to accessibility in graphical user interfaces (GUIs) by using deep learning models to predict natural-language labels for HTML elements, enabling better navigation for visually impaired users \cite{DBLP:conf/emnlp/LiLHZLG20, DBLP:conf/icse/ChenCXXZ0W20}.

\emph{Semi-automatic methods for Correcting Accessibility Violations } suggest an initial correction to a human expert for refinement. \cite{DBLP:conf/ccnc/MangiatordiL18, DBLP:conf/hcomp/SalisburyKM17, DBLP:journals/taccess/PereiraGRGD24} proposed using image recognition techniques to provide a first draft of the alt text, followed by human validation. 

\emph{LLM Prompting for Correcting  Accessibility Violations.} \label{relatedWork} LLMs have been shown to effectively generate code and detect errors \cite{DBLP:conf/icse/NamMHVM24, DBLP:conf/nips/LiuXW023, DBLP:conf/chi/Vaithilingam0G22, DBLP:conf/w4a/AljedaaniHAEF24, DBLP:journals/corr/abs-2107-03374, DBLP:journals/corr/abs-2108-07732, DBLP:conf/naacl/AhmadCRC21}. Recent studies have also shown that LLMs can generate accessibility-conformant code \cite{DBLP:conf/w4a/AljedaaniHAEF24, iyad2024}. Othman et al. \cite{DBLP:conf/petra/OthmanDJ23} used contextual prompting, where ChatGPT was provided with a non-conformant accessibility code and WCAG 2.1 guidelines to generate corrections. Delnevo et al. \cite{DBLP:conf/ccnc/DelnevoAM24} explored zero-shot prompting to correct accessibility violations by instructing ChatGPT to determine whether HTML elements are accessibility complaints. Huang et al. \cite{DBLP:journals/corr/abs-2401-16450} investigated three prompting techniques: Reasoning + Acting (ReAct), Few-Shot, and Chain of Thought (CoT).  ReAct, which combines reasoning steps with interactive responses, achieved the best performance on their dataset.
We adopt these three prompting techniques—contextual \cite{DBLP:conf/petra/OthmanDJ23}, zero-shot \cite{DBLP:conf/ccnc/DelnevoAM24}, and ReAct \cite{DBLP:journals/corr/abs-2401-16450}—as baselines for comparison against our proposed method. The prompt templates used for each baseline are directly adapted from the original works and are provided in Table \ref{tab:baseline_prompt_templates}. While these methods were effective in correcting some syntax and layout accessibility violations, they don't consider correcting semantic accessibility violations. Additionally, the lack of comparison with human-generated corrections limits a comprehensive evaluation of how well these models handle more complex, real-world semantic accessibility challenges.

\emph{LLMs as Assistive Agents for Web Accessibility. } LLMs can act as interactive assistive agents, helping users navigate and operate inaccessible Web pages.
Kodandaram et al. \cite{kodandaram2024enabling} introduced a system that uses LLMs to interpret screen elements and execute spoken user commands such as clicking buttons or filling out forms. Similarly, Mehendale and Walishetti \cite{mehendale2024dexassist} proposed the use of one LLM to translate voice input into Web interactions and the use of a second LLM to validate the outcome, thus improving the success rates of task execution. While such systems offer compelling advantages, the translation from content to accessible representation must be performed repeatedly for every user interaction, incurring high computational costs. Furthermore, because the transformations occur in real-time and do not persist in the Web content itself, they are difficult to monitor and correct. Their lack of transparency hinders reproducibility and quality assurance.

\subsection{Prompt Engineering Techniques}

LLMs are highly sensitive to prompt structure, producing more accurate and contextually relevant responses when guided by well-structured prompts. This observation has given rise to the field of prompt engineering, which involves designing instructions that generative AI models can effectively interpret \cite{DBLP:journals/corr/abs-2302-11382, DBLP:journals/corr/abs-2402-07927}. We survey techniques that we have adopted in the following subsections.

\subsubsection{Role-play prompting} directs LLMs to adopt specific personas, characters, professions, or roles (e.g., doctor, teacher, or historical figure), priming them to provide responses that align with the expert knowledge of that role. This approach can help guide the model's output to be more precise and factually accurate without altering the underlying capabilities of the LLM, as evidenced by studies \cite{roleplay, DBLP:conf/acl/WangPQLZWGGN00024, DBLP:journals/nature/ShanahanMR23, fathallah2024neon, DBLP:journals/corr/abs-2406-18460}.

\subsubsection{Contextual prompting } enriches prompts with task-specific information to guide the model's response \cite{DBLP:journals/corr/abs-2111-02643,DBLP:conf/cvpr/RaoZ0TZH0L22}. Dynamic contextual prompting adapts the context to each task, improving accuracy and reducing hallucinations. Prior work \cite{DBLP:journals/corr/abs-2111-02643,DBLP:conf/cvpr/RaoZ0TZH0L22} shows that task-specific context yields more relevant and higher-quality outputs than zero-shot.

\subsubsection{Metacognitive Prompting, } metacognition refers to the ability to self-reflect and critically evaluate one’s own cognitive processes \cite{frith2012role,fleming2014measure}. Metacognitive prompting \cite{DBLP:conf/www/0002LJND24,DBLP:journals/corr/abs-2308-05342} mimics human cognitive steps through the following stages: \textbf{(a) Self-understanding:} the model assesses its understanding of the prompt by identifying and interpreting relevant information, \textbf{(b) Reflection:} the model undergoes preliminary judgment, forming an initial response, followed by inference evaluation, where it reflects on and refines its initial interpretation, and \textbf{(c) Self-regulation:} the model finalizes its response and performs a confidence assessment to evaluate the reliability of its decision.

\subsubsection{Corrective Re-prompting} involves re-querying the model with feedback when its response fails to meet specific conditions \cite{DBLP:journals/corr/abs-2211-09935}. This feedback explains the failure (e.g., unmet preconditions), helping the model generate a more suitable response.

\thispagestyle{nofooter}
\noindent
\begin{minipage}[t]{0.50\textwidth}
{\small
\begin{lstlisting}[basicstyle=\scriptsize\ttfamily, caption={Synthesized HTML Web page from our dataset containing accessibility violations. The listing shows the accessibility violations detected by \method.}, numbers=left, numberstyle=\tiny, stepnumber=1, numbersep=5pt, label={lst:accessibility_not_fixed}]
<!DOCTYPE html>
<!-- Violation 1: Missing Language Attribute (Syntax)-->
<html>
<head>
<!--Violation 2: Viewport prevents scaling, zooming (Layout)-->
    <meta name="viewport" content="width=device-width,minimum-scale=1,maximum-scale=1,user-scalable=no">
    <title>Health Information Portal</title>
    <link rel="stylesheet" href="styles.css">
</head>
<body style="font-family: Arial, sans-serif; background-color: #fafafa; color: #888888;">
<!-- Violation 3: Contrast between foreground and background doesn't meet WCAG 2 AA minimum contrast ratio thresholds (Layout)-->
    <div style="background-color: #333; padding: 15px;">
    <h1 style="color: #888888;">Welcome to the Vitamin Resource Hub Portal</h1>
    <div class="content">
    <h2>Learn how vitamins can boost your health and well-being</h2>
    <p>Essential Vitamins</p>
<!-- Violation 4: Scrollable region not keyboard accessible (Syntax)-->
    <div class="scrollable-content">     
        <p>Vitamin A </p>
        <p>Vitamin C</p>
        <p>Vitamin D</p>
        <p>Vitamin E</p>
        <p>Vitamin K</p>
    </div>
<!-- Violation 5: Image alt text not descriptive (Semantic)-->
    <img src="https://img.Webmd.com/woman.jpg" alt="image">
    <!-- Violation 6: Table structure missing accessibility attributes (Syntax)-->
    <table>
        <tr>
            <td>Vitamin</td>
            <td>Recommended Daily Amount</td>
        </tr>
        <tr>
            <td>Vitamin C</td>
            <td>75 mg</td>
        </tr>
        <tr>
            <td>Vitamin D</td>
            <td>600 IU</td>
        </tr>
    </table>
<!-- Violation 7: Link without discernible text (Syntax)-->
    <a href="vitamin-guide.pdf"><img src="https://vitaminguide.png" height="95" width="95"></a>
    <p>Select which topics you want to include in your personalized vitamin guide:</p>
<!-- Violation 8: Nested interactive elements (Syntax)-->
    <div>
        <label class="vitamin-checkbox" role="checkbox" aria-checked="false">
            <input type="checkbox"> Special Vitamin Tips for Kids
        </label>
        <label class="vitamin-checkbox" role="checkbox" aria-checked="false">
            <input type="checkbox"> Vitamins for Skin Health
        </label>
        <label class="vitamin-checkbox" role="checkbox" aria-checked="false">
            <input type="checkbox"> Daily Multivitamin Recommendation
        </label>
        </div>
<!-- Violation 9: Button without discernible text (Syntax)-->
    <button class="subscribe-button" type="button">
    <img src="https://ScreenShot.png">
    </button> </div> </div> </body></html>
\end{lstlisting} }
\end{minipage}
\newpage
\hspace{0.04\textwidth} 
\begin{minipage}[t]{0.50\textwidth}
{\small
\begin{lstlisting}[basicstyle=\scriptsize\ttfamily, caption={\method \space delivers this accessibility-compliant HTML code when provided with Listing~\ref{lst:accessibility_not_fixed}}., numbers=left, numberstyle=\tiny, stepnumber=1, numbersep=5pt, label={lst:accessibility_fixed}]
<!DOCTYPE html>
<!-- Fixed Violation 1: Added Language Attribute -->
<html lang="en">
<head>
<!-- Fixed Violation 2: Viewport allow scaling, zooming-->
    <meta name="viewport" content="width=device-width, initial-scale=1">
    <title>Health Information Portal</title>
    <link rel="stylesheet" href="styles.css">
</head>
<body style="font-family: Arial, sans-serif; background-color: #fafafa; color: #fafafa;">
<!-- Fixed Violation 3: Changed contrast between foreground and background to meet WCAG 2 AA standards -->
    <div style="background-color: #333; padding: 15px;">
    <h1 style="color: #fafafa;">Welcome to the Vitamin Resource Hub Portal</h1>
    <div class="content">
    <h2>Learn how vitamins can boost your health and well-being</h2>
    <p>Essential Vitamins</p>
<!-- Fixed Violation 4: Scrollable region accessible via keyboard (added tabindex attribute) -->
    <div class="scrollable-content" tabindex="0" >     
        <p>Vitamin A </p>
        <p>Vitamin C</p>
        <p>Vitamin D</p>
        <p>Vitamin E</p>
        <p>Vitamin K</p>
    </div>
<!-- Fixed Violation 5: Replaced nondescriptive alt text--> <img src="https://img.Webmd.com/woman.jpg" alt=" Woman reading vitamin box label while standing in a pharmacy aisle.">
<!-- Fixed Violation 6: Table structure & accessibility attributes (<th>) -->
    <table>
    <thead>
        <tr>
            <th scope="col">Vitamin</th>
            <th scope="col">Recommended Daily Amount</th>
        </tr> </thead> <tbody>
        <tr>
            <td>Vitamin C</td>
            <td>75 mg</td>
        </tr>
        <tr>
            <td>Vitamin D</td>
            <td>600 IU</td>
        </tr> </tbody> </table>
<!-- Fixed Violation 7: Added accessible text for the link-->
    <a href="vitamin-guide.pdf"><img src="https://vitaminguide.png" alt="Link to download  PDF guide on vitamins"></a>
    <p>Select which topics you want to include in your personalized vitamin guide:</p>
<!-- Fixed Violation 8: Removed improper nesting of interactive elements -->
    <div>
        <label class="vitamin-checkbox">
            <input type="checkbox" aria-label="Special Vitamin Tips for Kids"> Special Vitamin Tips for Kids
        </label>
        <label class="vitamin-checkbox">
            <input type="checkbox" aria-label="Vitamins for Skin Health"> Vitamins for Skin Health
        </label>
        <label class="vitamin-checkbox">
            <input type="checkbox" aria-label="Daily Multivitamin Recommendation"> Daily Multivitamin Recommendation
        </label>
    </div>
<!-- Fixed Violation 9: Added discernible text for button -->
    <button class="subscribe" type="button" aria-label="Subscribe to Vitamin Newsletter">
    <img src="https://ScreenShot.png" alt="Subscribe Button">
    </button> </div> </div> </body></html>
\end{lstlisting} }
\end{minipage}

\section{Our Taxonomy of Web Accessibility Violations} \label{sec:taxonomy}
We introduce a structured taxonomy of Web accessibility violations that categorizes them into three distinct dimensions: \emph{Syntactic}, \emph{Semantic}, and \emph{Layout}. To construct the taxonomy, we adopted a methodology from prior work on taxonomy standardization and classification~\cite{vsmite2014empirically, usman2017taxonomies}. 
In the first stage, we conducted a literature review to identify recurring Web accessibility violations, including studies incorporating user feedback. 

In the second stage, we analyzed WCAG 2.1 alongside rule definitions from automated testing tools (Axe-Playwright~\cite{PlaywrightAPI}, WAVE \cite{Webaim2025wave}, AChecker \cite{achecker2024}), extracting and consolidating violation types.  We adopt naming conventions from Axe-Playwright to ensure consistency and interpretability (e.g., \texttt{button-name}, \texttt{html-has-lang}, \texttt{color-contrast}). Each type was defined by its associated WCAG and required correction logic. For instance, although both \texttt{form-label-mismatch} and \\ \texttt{button-label-mismatch} relate to label mismatch, we identified them as distinct semantic accessibility violations: the former refers to form elements whose labels fail to describe their input purpose, while the latter involves buttons with labels that do not reflect their action. Violations exhibiting functional overlap were merged or redefined to ensure mutually exclusive boundaries.

In the third stage, we validated the taxonomy’s relevance and generalizability by evaluating its alignment with outputs from automated testing tools (Axe-Playwright~\cite{PlaywrightAPI}, WAVE \cite{Webaim2025wave}, AChecker \cite{achecker2024}) and cross-referencing the prevalence of accessibility violations against large-scale reports from the WebAIM 2025 study \cite{Webaim2025}. For semantic accessibility violations— absent from automated tools—we conducted manual audits of Web pages from \cite{Webaim2025}. Violations were retained only if observed repeatedly across Web pages. 

\begin{figure*}
  \centering
  \caption{Interface of the Web page before (a) and after (b) correction with \textit{AccessGuru}, corresponding to the code in Listing~\ref{lst:accessibility_not_fixed} and Listing~\ref{lst:accessibility_fixed}. This example includes corrections of all three accessibility violation types: (1) \emph{Syntactic}—added missing table headers and ARIA attributes, (2) \emph{Semantic}—replaced non-descriptive alt text with meaningful descriptions, and (3) \emph{Layout}—adjusted color values to improve contrast. The overall visual appearance remains largely unchanged for typical users.}
  \Description{Interface of the Web page before (a) and after (b) correction with AccessGuru, corresponding to the code in Listing 1 and Listing 2. This example includes corrections of all three violation types: (1) Syntactic, added missing table headers and ARIA attributes; (2) Semantic, replaced non-descriptive alt text with meaningful descriptions; and (3) Layout, adjusted color values to improve contrast. The overall visual appearance remains largely unchanged for typical users.}
  \includegraphics[width=15.5cm, alt={Interface of the Web page before (a) and after (b) correction with AccessGuru, corresponding to the code in Listing~\ref{lst:accessibility_not_fixed} and Listing~\ref{lst:accessibility_fixed}. This example includes corrections of all three violation types: (1) Syntactic, added missing table headers and ARIA attributes; (2) Semantic, replaced non-descriptive alt text with meaningful descriptions; and (3) Layout, adjusted color values to improve contrast. The overall visual appearance remains largely unchanged for typical users.}]{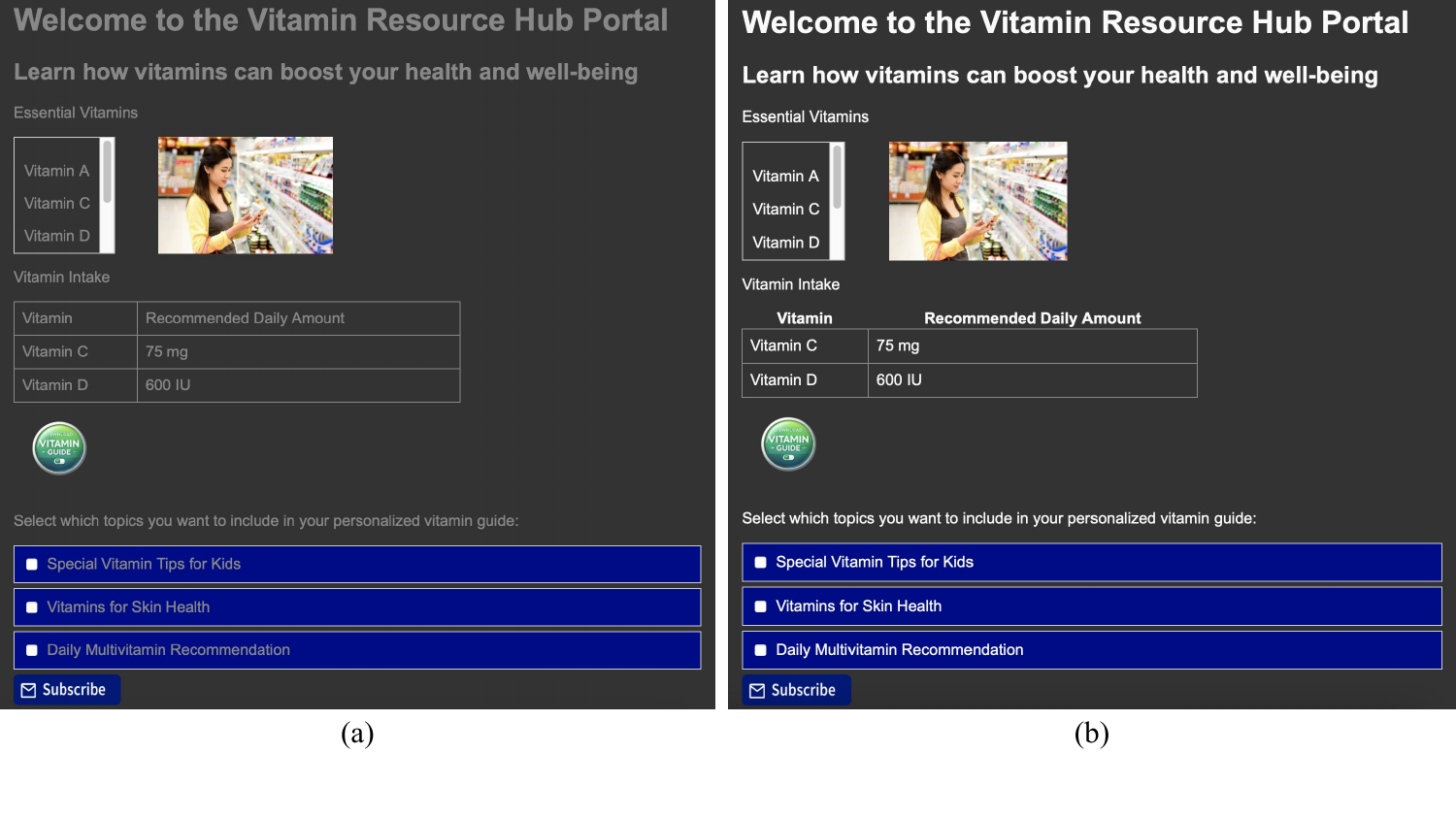}

  \label{fig:ui}
\end{figure*}

\begin{itemize}
    \item \emph{Syntactic accessibility violations} arise when the required accessibility-enhancing HTML elements and attributes are missing or malformed. These include elements like alt text for images or ARIA (Accessible Rich Internet Applications) attributes, which help define the behavior and purpose of interactive components. Syntactic accessibility compliance is fulfilled if all required and useful syntactic constructs for indicating accessibility information are present. For instance, in Listing \ref{lst:accessibility_not_fixed} Violation 6 (line 27), a table presenting vitamin data lacks accessibility syntax elements like \texttt{<th>} and scope attributes. This omission hinders screen readers from correctly identifying table headers, making it difficult for visually impaired users to understand the relationship between vitamins and their recommended daily amounts.

\item \emph{Layout accessibility violations} occur when the visual and spatial arrangement of content fails to meet accessibility guidelines. This includes sufficient color contrast between text and background to ensure readability. It also includes ensuring that users can adjust content presentation based on their needs, such as enabling text scaling and zooming.

\item \emph{Semantic accessibility violations} occur when HTML accessibility elements are present but fail to convey meaningful content. In Listing~\ref{lst:accessibility_not_fixed} Violation 5 (line 25), an image includes an alt attribute set to the generic string \texttt{"image"}—a common issue caused by auto-generated text. The corrected version is shown in Listing~\ref{lst:accessibility_fixed}, line 25.
\end{itemize}
Our taxonomy comprises over 112 violation types. Each violation is annotated with associated WCAG guidelines, impact level, and required supplementary information (e.g., \\ \texttt{image-alt-not-descriptive} requires the image alongside the HTML to detect and correct the violation). For syntactic and layout accessibility violations, impact levels are derived from existing tools \cite{PlaywrightAPI}, which judge the severity of each violation based on how much it hinders user interaction with the Web content. For semantic accessibility violations, we manually assign impact levels using analogous criteria (e.g., \texttt{button-label-mismatch} inherits the "Critical" level from its syntactic counterpart \texttt{button-name}). A representative subset is shown in Table~\ref{tab:important_violations}, and the complete taxonomy can be found in the supplementary material of this paper.

\section{Methodology}
\method \space addresses Web accessibility violations in two stages: first, by automatically detecting accessibility violations, and second, by automatically generating corrections for the detected accessibility violations. 
\subsection{\method \emph{Detect}: Web Accessibility Violation Detection} \label{detect}

\begin{figure*}
  \centering
    \caption{ Overview of the \method \emph{Detect}. Given a raw HTML document (left), it applies two detectors: (1) a syntax and layout detector based on the Axe-Playwright accessibility testing engine and (2) an LLM-based semantic detector. The output set of detected accessibility violations (right).}
    \Description{{Overview of the AccessGuru Detection Module. Given a raw HTML document (left), the detection module applies two detectors: (1) a syntax and layout detector based on the Axe-Playwright accessibility testing engine and (2) an LLM-based semantic detector. The module outputs a set of detected accessibility violations (right).}}
  \includegraphics[width=18cm, alt={Overview of the AccessGuru Detection Module. Given a raw HTML document (left), the detection module applies two detectors: (1) a syntax and layout detector based on the Axe-Playwright accessibility testing engine and (2) an LLM-based semantic detector. The module outputs a set of detected accessibility violations (right).}]{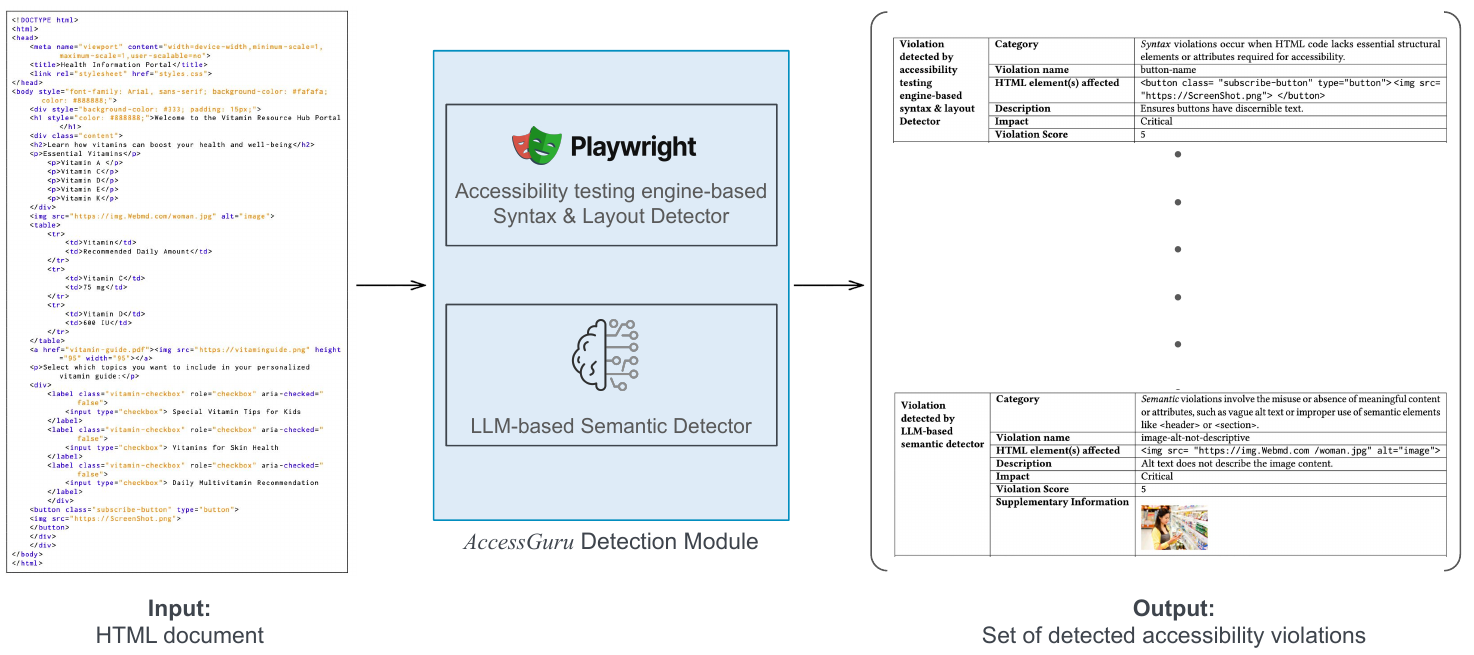}
  \label{detectFigure}
\end{figure*}

\method \emph{Detect} uses an Axe-Playwright-based detector to detect syntactic and layout accessibility violations following methodologies from similar studies \cite{DBLP:journals/corr/abs-2401-16450,acosta2020dataset} and an LLM to detect semantic accessibility violations (See Figure~\ref{detectFigure}). Both detectors are guided by our taxonomy, which defines violation types, required context (e.g., rendered images for evaluating \texttt{image-alt-not-descriptive}), and severity levels.

Given a target HTML document, \method \emph{Detect} executes the following steps:
\begin{enumerate}   
    \item The Axe-Playwright tool is executed on the target HTML document; the tool generates a detailed violation report, which includes for each violation: \textbf{(a)} violation name, \textbf{(b)} affected HTML elements, \textbf{(c)} violation description,  and \textbf{(d)} impact level. We enrich the violation with \textbf{(e)} numerical violation scores ranging from 1 (lowest) to 5 (highest) by mapping the qualitative impact levels reported by Axe-Playwright ("cosmetic", "minor", "moderate", "serious", and "critical") to corresponding numeric values.

    \item For each violation, we use the violation name to query our predefined taxonomy, which specifies whether supplementary information beyond the HTML is required for correction (see Table~\ref{tab:important_violations}). If additional data is needed—such as color values for contrast violations—we extract it as \textbf{(f)} supplementary information from the rendered HTML document (e.g., computed CSS styles for background and foreground colors). 
\end{enumerate}

To detect semantic accessibility violations, \method\ uses an LLM-based semantic detector that runs in parallel to the Axe-Playwright-based detector. Given the raw HTML document and a screenshot of its rendered view.  \method \emph{Detect} executes the following steps:
\begin{enumerate}
\item The HTML document is rendered in a browser that captures a full-page screenshot to account for scrolling behavior. The screenshot is taken at a fixed viewport width of 1440 pixels and a height equal to the full scrollable length of the page, resulting in an image with a height up to 1440 times the viewport height, depending on page length.

\item The LLM is prompted with the HTML document, screenshot, and our semantic violation taxonomy. We show the prompt template in Table \ref{tab:semantic_detection_prompt}. The prompt instructs the LLM to identify all semantic accessibility violations and enclose violation \textbf{(a)} names and \textbf{(b)} affected HTML elements within string markers (e.g., \texttt{[START]}, \texttt{[END]}). To support reasoning over non-textual content (Web page screenshot), the LLM must exhibit multimodal capabilities.
  \item Each marked segment by the LLM is matched against the original HTML code to ensure it refers to an existing element. 
\item If any of the returned HTML segments by the LLM do not match an element in the original HTML, it is discarded as hallucinations and excluded from the final set of detected violations.
  \item For each violation, we use the violation name returned by the LLM to look up our taxonomy to assign \textbf{(c)} violation description, \textbf{(d)} impact level ("cosmetic", "minor", "moderate", "serious", and "critical"). We enrich the violation with \textbf{(e)} numerical violation scores ranging from 1 (lowest) to 5 (highest).

\item  For each violation, we use the violation name to look up our taxonomy to determine if supplementary information to the HTML is required (e.g., images for \texttt{image-alt-not-descriptive}, videos for \\ \texttt{video-caption-alt-not-descriptive}). If so, we extract \textbf{(f)} supplementary information crucial for correcting accessibility violations. 
\end{enumerate}
For illustration purposes, Listing~\ref{lst:accessibility_not_fixed} includes annotated comments to indicate the locations of accessibility violations; these comments are added manually for reader clarity and are not provided as input to \method \emph{Detect}. When the unannotated HTML in Listing~\ref{lst:accessibility_not_fixed} is processed, \method \emph{Detect} identifies all the annotated accessibility violations. Each detector runs independently, and the aggregated output is stored as a single JSON file containing a set of violation entries. Samples from the output are shown in Tables \ref{sample} and \ref{semanticExample}.

\method \emph{Detect} outputs a set of detected Web accessibility violations $V = \{v_1, v_2, \ldots, v_n\}$, where each $v_i$ represents an individual violation (see Figure~\ref{detectFigure}).

\subsection{\method \emph{Correct}: \space Web Accessibility Violation Correction} \label{correctt}

\method \emph{Correct} operates on the output of the \method \emph{Detect} —a set of detected Web accessibility violations—taking as input $V = \{v_1, v_2, \ldots, v_n\}$, where each $v_i$ represents an individual We accessibility violation (see Figure~\ref{detectFigure}).

\begin{figure*}
  \centering
    \caption{Overview of \method \emph{Correct} in: For each Web accessibility violation detected by \method \emph{Detect} (examples of the input accessibility violations are shown in Tables \ref{sample} and \ref{semanticExample}), the LLM is prompted to generate the corrected code. The generated code is assigned a violation score; if the violation score remains above zero, corrective re-prompting is applied to improve the response further.}
       \Description{Overview of AccessGuru's correction module: For each Web accessibility violation $v$ detected by AccessGuru's Detection module, the LLM is prompted to generate the corrected code. The generated code is assigned a violation score; if the violation score remains above zero, corrective re-prompting is applied to improve the response further.}
  \includegraphics[width=\textwidth, alt={Overview of AccessGuru's correction module: For each Web accessibility violation $v$ detected by \method \emph{Detect}, the LLM is prompted to generate the corrected code. The generated code is assigned a violation score; if the violation score remains above zero, corrective re-prompting is applied to improve the response further.}]{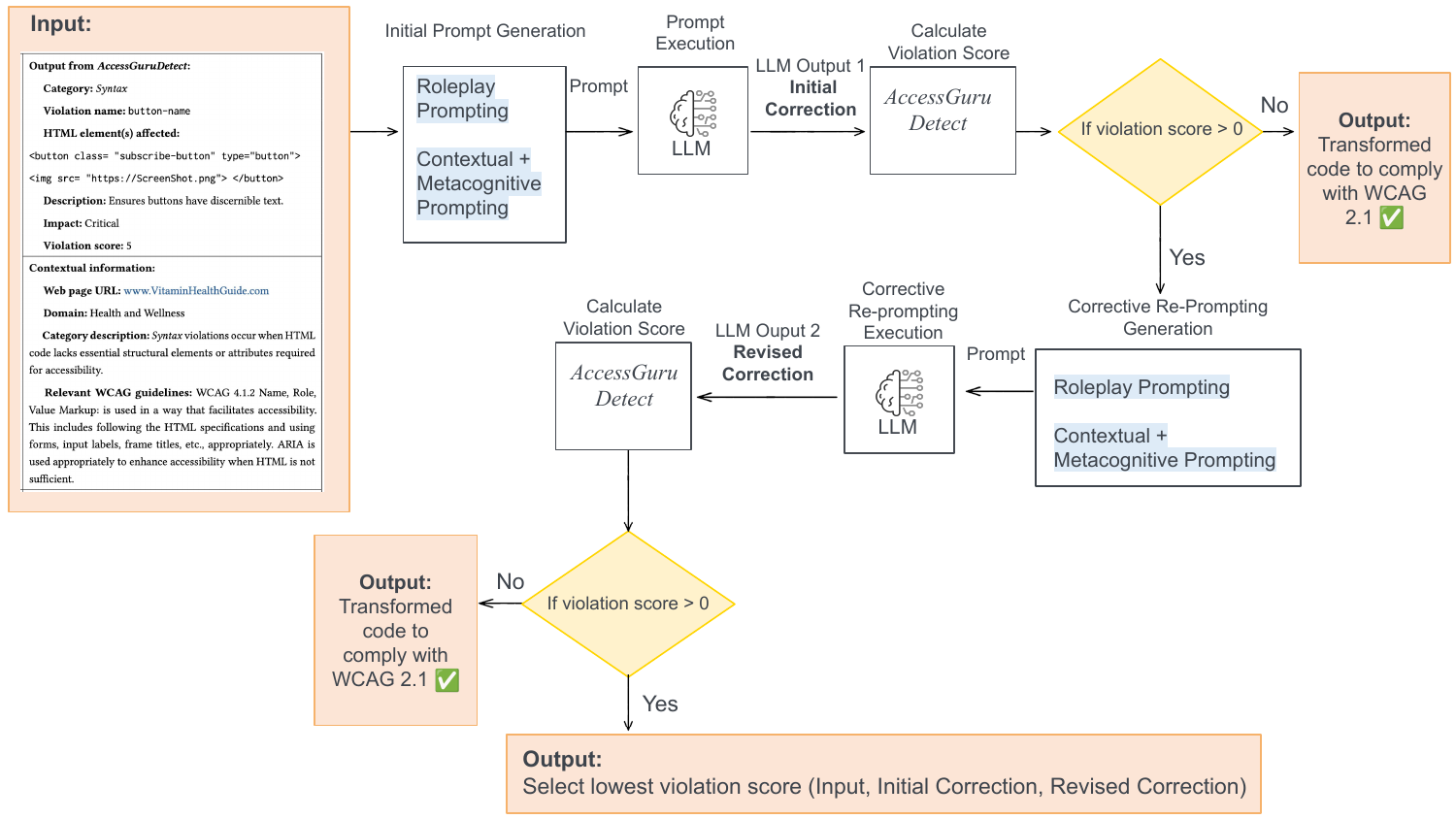}
  \label{fig:overview}
\end{figure*}

Each violation includes the affected HTML element(s), violation metadata (See Tables \ref{sample} and \ref{semanticExample} for sample inputs). The goal is to generate correct HTML for each violation that minimizes the total violation score, as shown in Figure~\ref{fig:overview} and detailed in Algorithm~\ref{alg:correction}.

For each violation $v$, detected by \method \emph{Detect} and enriched with violation-specific data such as supplementary information when required (e.g., images for \\ \texttt{image-alt-not-descriptive}, videos for \\ \texttt{video-caption-alt-not-descriptive}), we construct an \emph{initial prompt} (Algorithm~\ref{alg:correction}: line 3) and submit it to a pre-trained LLM (Algorithm~\ref{alg:correction}: line 4). For semantic violations, the LLM must possess multimodal capabilities to reason over non-textual content (e.g., images for \texttt{image-alt-not-descriptive}, videos for \texttt{video-caption-alt-not-descriptive}).

The \emph{initial prompt} integrates three techniques: (1) \textbf{role-play prompting}, (2) \textbf{contextual prompting}, and (3) \textbf{metacognitive prompting}. The role-play technique embeds a Web accessibility expert persona (See Table~\ref{persona}) to guide the LLM toward expert-like behavior. Contextual prompting enriches the prompt with violation-specific data— Web page URL, domain, WCAG guidelines, and a violation category description. Driven by our taxonomy, which defines the categories (\emph{Syntactic}, \emph{Semantic}, \emph{Layout}) and their associated characteristics, we include the description of the violation category in each prompt. For example, layout violations may require CSS and responsive design awareness while preserving visual integrity for all users. These components are integrated using metacognitive prompting; we show the prompt template Table~\ref{fig:mp}.

The LLM returns a correction ($LLMOutput1$), enclosed between string markers (e.g., \texttt{\#\#\#START\#\#\#} and \texttt{\#\#\#END\#\#\#}), as specified in the prompt. We extract the HTML snippet using regular expressions and evaluate it with \method \emph{Detect} to assign a violation score (Algorithm~\ref{alg:correction} line 5). If the violation score is zero, the correction is accepted (Algorithm~\ref{alg:correction}: line 6). Otherwise, we apply corrective re-prompting by adding feedback and resubmitting the prompt (Algorithm~\ref{alg:correction}: lines 8–10), yielding a revised correction ($LLMOutput2$); we show the prompt template Table~\ref{fig:mp_rePrompt}.

If the revised output still contains accessibility violations (Algorithm~\ref{alg:correction}: line 11), we compare the scores of $v$, $LLMOutput1$, and $LLMOutput2$ (Algorithm~\ref{alg:correction}: line 12). The version with the lowest score is chosen as the final correction (Algorithm~\ref{alg:correction}: line 13), with ties resolved by preferring the most recent output.

Our scoring mechanism is robust to new LLM-induced accessibility violations by treating all accessibility violations equally in the total violation score. For instance, if $LLMOutput1$ yields a violation score of 4, while $LLMOutput2$ introduces two accessibility violations scoring 1 and 2 (total 3), we prefer $LLMOutput2$. Conversely, if both LLM outputs score worse than the original input (e.g., $LLMOutput1$ violations score = 4 and $LLMOutput2$ violations score =  3 vs. $v$ violations score = 2), we select the original code as the output.

The output is the correction generated by \method \emph{Correct}, which is appended to its corresponding JSON entry in the set, as shown in Tables \ref{sample} and \ref{semanticExample}. Resulting in a final JSON file that contains all detected accessibility violations along with their suggested corrections.

\begin{algorithm}

\caption{\method \emph{Correct}}
\label{alg:correction}

\begin{algorithmic}[1]
\State \textbf{input:} Set of detected accessibility violations $V = \{v_1, v_2, \ldots, v_n\}$
\State \textbf{ouput:} Set of corrected HTML segments $\{c_1, c_2, \ldots, c_n\}$ 
\For{each violation $v \in V$}
    \State \textbf{Construct} \emph{Initial Prompt}($v$)
    \State \textbf{Submit} prompt to LLM to obtain $LLMOutput1$
    \If{$ViolationScore(LLMOutput1) = 0$}
        \State $c \gets LLMOutput1$
    \Else
        \State \textbf{Construct} \emph{Corrective Prompt}($v$)
        \State \textbf{Submit} prompt to LLM to obtain $LLMOutput2$
        \If{$ViolationScore(LLMOutput2) = 0$}
            \State $c \gets LLMOutput2$
        \Else
            \State $c \gets SelectBest(v, LLMOutput1, LLMOutput2)$
        \EndIf
    \EndIf
\EndFor
\State \Return $\{c_1, c_2, \ldots, c_n\}$
\end{algorithmic}
\end{algorithm}
\newpage
\subsubsection{Correction Independence and Overwrites: }

\method\emph{Correct}{} applies corrections sequentially, generating a separate correction for each violation. Our taxonomy ensures that violations are non-overlapping, so each correction typically modifies a different attribute of the same HTML node.  Consider the following example, where a single HTML node contains two violations: \\
\noindent\texttt{<a href="subscribe.html" style="color:\#767676; \\ background:\#303030;">} \\
\texttt{\ \ <img src="sub.png">} \texttt{</a>} \\
\noindent\method{} generates two independent corrections:
\begin{itemize}
\item \textbf{Violation A: \texttt{link-name}}  \\
\texttt{- <a href="subscribe.html" ...>} \\
\texttt{+ <a href="subscribe.html" aria-label="Subscribe" ...>}
\item \textbf{Violation B: \texttt{color-contrast}} \\ 
\texttt{- style="color:\#767676;background:\#303030;"} \\
\texttt{+ style="color:\#FFFFFF;background:\#1A1A1A;"}
\end{itemize}
Each correction targets a distinct part of the node and can be applied independently. However, rare edge cases may result in overlapping edits. For example:\\
\texttt{<html lang="eng" xml:lang="en-GB">} \\
\noindent\method{} suggests:
\begin{itemize}
    \item \textbf{Violation A: \texttt{html-lang-valid}}  \\
\texttt{- lang="eng"} \\
\texttt{+ lang="en"}
\item \textbf{Violation B: \texttt{html-xml-lang-mismatch}}\\
\texttt{- lang="eng"} \\
\texttt{+ lang="en-GB"}
\end{itemize}
The second correction may overwrite the first; any overwrite, therefore, further refines the correction.

\renewcommand{\arraystretch}{1.3} 
\begin{table*}[t]
\caption{Example JSON entry from \method’s output for a syntax violation from the HTML document in Listing~\ref{lst:accessibility_not_fixed} (Violation 9, Line 57). It consists of: (1) the violation $v$ detected by \method{} \emph{Detect}, including metadata; (2) contextual information; and (3) the correction generated by \method{} \emph{Correct}.}
\label{sample}

\centering
\resizebox{\textwidth}{!}{
\begin{tabular}{|p{3cm}|p{15cm}|}
\hline
\multirow{10}{=}{\textbf{Input to \method{} \emph{Correct}}} 
& \textbf{Output from \method \emph{Detect}: } \\ 
& \hspace*{1em} \textbf{Category:} \textit{Syntax} \\ 
& \hspace*{1em} \textbf{Violation name:} \texttt{button-name} \\ 
& \hspace*{1em} \textbf{HTML element(s) affected:} \\ 
& \hspace*{1em}  \texttt{<button class="subscribe-button" type="button"> <img src="https://ScreenShot.png"> </button>} \\ 
& \hspace*{1em} \textbf{Description:} Ensures buttons have discernible text. \\ 
& \hspace*{1em} \textbf{Impact:} Critical \\ 
& \hspace*{1em} \textbf{Violation score:} 5 \\ \cline{2-2}
& \textbf{Contextual information:} \\ 
& \hspace*{1em} \textbf{Web page URL:} \url{www.VitaminHealthGuide.com} \\ 
& \hspace*{1em} \textbf{Domain:} Health and Wellness \\ 
& \hspace*{1em} \textbf{Category description:} \textit{Syntax} violations occur when HTML code lacks essential structural elements or attributes required for accessibility. \\ 
& \hspace*{1em} \textbf{Relevant WCAG guidelines:} WCAG 4.1.2 Name, Role, Value Markup: is used in a way that facilitates accessibility. This includes following the HTML specifications and using forms, input labels, frame titles, etc., appropriately. ARIA is used appropriately to enhance accessibility when HTML is not sufficient. \\ \hline

\textbf{Output from \method{} \emph{Correct}} 
& \texttt{<button class="subscribe" type="button" aria-label="Subscribe to Vitamin Newsletter"> <img src="https://ScreenShot.png" alt="Subscribe Button Image"> </button>} \\ \hline
\end{tabular}}
\end{table*}

\renewcommand{\arraystretch}{1.3}
\begin{table*}
\caption{Example JSON entry from \method’s output for a semantic violation from Listing~\ref{lst:accessibility_not_fixed} (Violation 5, Line 25). It consists of: (1) the violation $v$ output from \method{} \emph{Detect}; (2) input to \method \emph{Correct}, which is the output from \method{} \emph{Detect} along with the contextual information; and (3) the output correction generated by \method{} \emph{Correct}.}

\centering
\resizebox{\textwidth}{!}{
\begin{tabular}{|p{3cm}|p{15cm}|}
\hline
\multirow{10}{=}{\textbf{Input to \method{} \emph{Correct}}} 
& \textbf{Output from \method \emph{Detect}:} \\ 
& \hspace*{1em} \textbf{Category:} \textit{Semantic} \\ 
& \hspace*{1em} \textbf{Violation name:} \texttt{image-alt-not-descriptive} \\ 
& \hspace*{1em} \textbf{HTML element(s) affected:} \texttt{<img src="https://img.Webmd.com/woman.jpg" alt="image">} \\ 
& \hspace*{1em} \textbf{Description:} alt text does not describe the image content. \\ 
& \hspace*{1em} \textbf{Impact:} Critical \\ 
& \hspace*{1em} \textbf{Violation score:} 5 \\ 
& \hspace*{1em} \textbf{Supplementary information:} \raisebox{-0.5\height}{\includegraphics[width=2cm]{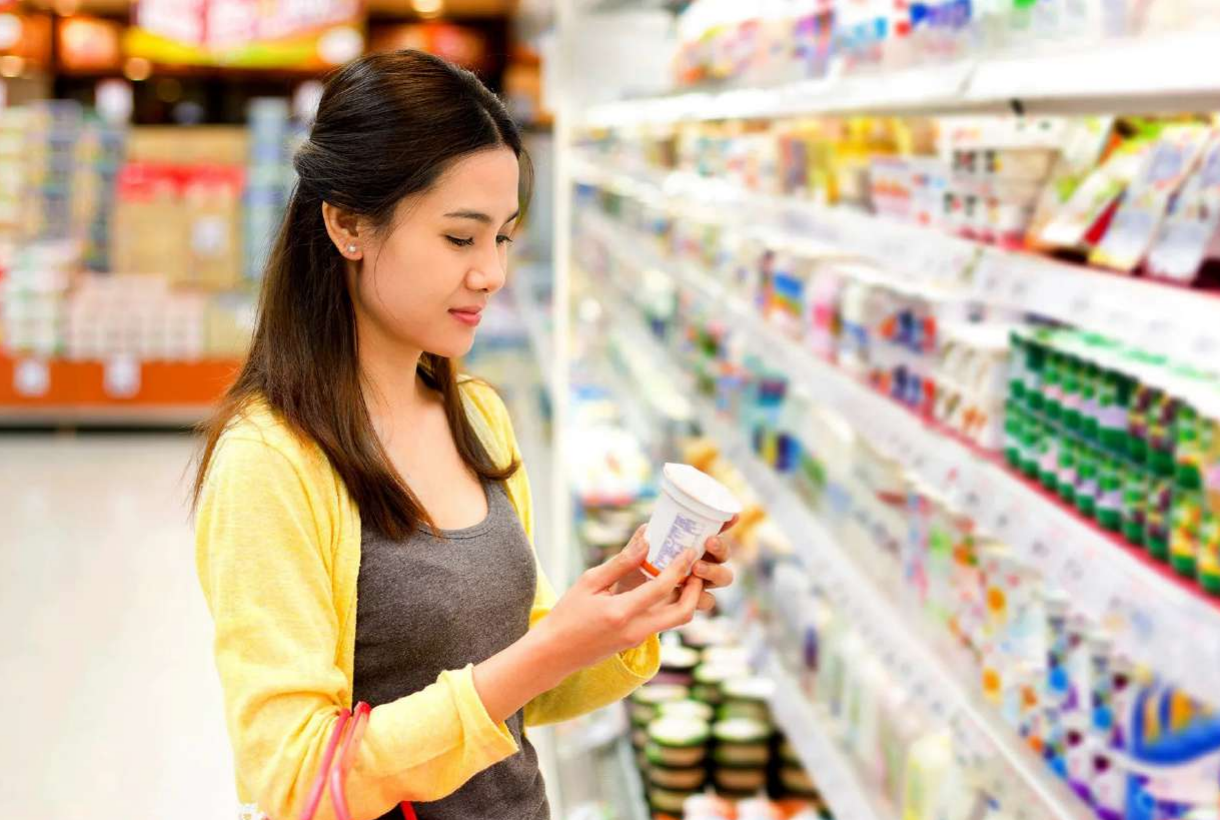}} \\ \cline{2-2}
& \textbf{Contextual information:} \\ 
& \hspace*{1em} \textbf{Web page URL:} \url{www.VitaminHealthGuide.com} \\ 
& \hspace*{1em} \textbf{Domain:} Health and Wellness \\ 
& \hspace*{1em} \textbf{Category description:} \textit{Semantic} violations involve the misuse or absence of meaningful content or attributes, such as vague alt text or improper use of semantic elements like \texttt{<header>} or \texttt{<section>}. \\ 
& \hspace*{1em} \textbf{Relevant WCAG guidelines:} WCAG 1.1.1 Non-text Content. All meaningful visual elements (e.g., images, image buttons, image maps) must have descriptive alternative text. Form controls, inputs, multimedia, and frames must include accessible names, labels, or titles to ensure clarity for assistive technologies. \\ \hline

\textbf{Output from \method{} \emph{Correct}} 
& \texttt{<img src="https://img.Webmd.com/woman.jpg" alt="Woman reading the label on a vitamin box while standing in a pharmacy aisle.">} \\ \hline
\end{tabular}}
\label{semanticExample}
\end{table*}

\section{Evaluation}
We evaluate the effectiveness of \method \space in detecting and correcting Web accessibility violations across the three categories defined in our taxonomy: \emph{syntactic}, \emph{semantic}, and \emph{layout}. Specifically, we ask: 

\begin{itemize}
\item Detection Evaluation: \textbf{RQ 1.} To what extent can our detection method identify accessibility violations across all three categories?
\item Syntactic and Layout Correction Evaluation: \textbf{RQ 2.} To what extent can the LLM generate HTML code that satisfies syntactic and layout accessibility compliance, effectively addressing Web accessibility violations?
    \item Semantic Correction Evaluation: \textbf{RQ 3.}  To what extent are the LLM-generated attributes semantically meaningful, as evaluated by human experts and in comparison to corrections made by human developers?
\end{itemize}

\subsection{Dataset of Web Accessibility Violations from \cite{DBLP:journals/corr/abs-2401-16450}}
We use the dataset from \cite{DBLP:journals/corr/abs-2401-16450} to evaluate whether \method \emph{Detect} can successfully identify the reported violations and assess whether \method \emph{Correct} can transform violations into accessibility-compliant HTML. The dataset consists of Web accessibility violations collected from 25 URLs, including popular Websites such as Google Calendar, Slack, and BBC. The dataset contains 171 rows of accessibility violations, which were identified using the Axe-Playwright and span 40 distinct accessibility violation types. Each violation is categorized by severity (ranging from cosmetic to critical) and includes details such as the Web page URL, violation name, description, and HTML elements.  

The dataset introduced by Huang et al.~\cite{DBLP:journals/corr/abs-2401-16450} represents a valuable step toward evaluating automated accessibility corrections. However, it remains limited in several important ways. The dataset is relatively small in scale and does not offer broad coverage of known Web accessibility violation types- 40 accessibility violation types. Moreover, upon manual inspection, we observed that the dataset provides only HTML for each violation, which is insufficient for certain accessibility violations. For example, addressing contrast violations requires access to the color values of foreground and background elements, which are often defined in external CSS and not reflected in the raw HTML. This omission restricts the ability to fully detect or correct such violations.

\subsection{Our novel dataset for benchmarking corrections of syntactic, layout, and semantic accessibility violations}

We introduce a new dataset that comprehensively covers syntactic, semantic, and layout accessibility violations, reflecting real-world accessibility violations. Our dataset was collected through a structured process guided by the WebAIM 2025 study \cite{Webaim2025}, which identifies commonly visited Websites with accessibility violations. To account for variation in accessibility violations across domains—e.g., government sites often use data tables, while e-commerce relies heavily on forms and images—we used GPT-4 \cite{DBLP:journals/corr/abs-2303-08774} to identify a diverse set of popular Web domains such as health, education, government, news, technology, and e-commerce, including multilingual Websites. We then asked GPT-4 to suggest representative URLs from each domain based on sites listed in the WebAIM 2025 study~\cite{Webaim2025}. This process yielded 448 URLs. 

We crawled each URL using Playwright and retained only those pages where \texttt{document.readyState === "complete"}.  We applied \method \emph{Detect} (Figure~\ref{detectFigure}) to each URL to identify Web accessibility violations.  This yielded 3,500 Web accessibility violations. The dataset spans over 112 distinct violation types across all three categories. To our knowledge, it is the most comprehensive publicly available dataset of real-world Web accessibility violations to date.  

To ensure our evaluation reflects real-world Web accessibility violations, we sampled a representative subset of 305 violations from our full dataset of 3,500 instances. Sampling aligns the distribution of violation types in the subset with real-world frequencies reported in the WebAIM 2025 study~\cite{Webaim2025}. This method avoids biases caused by overrepresented violation types in large-scale crawled data but does not reflect their actual prevalence across the Web. This subset size was chosen to enable controlled and consistent comparison across LLMs and baselines. Table~\ref{tab:dataset_distribution} compares the distribution of violation categories in our subset with that of the dataset from \cite{DBLP:journals/corr/abs-2401-16450}.
\renewcommand{\arraystretch}{1.2}
\begin{table}[h]
\centering
\caption{Distribution of accessibility violations in our dataset vs.\ \cite{DBLP:journals/corr/abs-2401-16450}}
\label{tab:dataset_distribution}
\small
\begin{tabular}{|l|c|c|}
\hline
\textbf{Violation Category} & \makecell{\textbf{\# in}  \textbf{Our Dataset}} & \makecell{\textbf{\# in}  \textbf{\cite{DBLP:journals/corr/abs-2401-16450}}} \\ \hline
Syntax   & 195 & 151 \\
Layout   & 55  & 20  \\ 
Semantic & 55  & 0   \\ \hline
\textbf{Total}     & \textbf{305} & \textbf{171} \\ \hline
\textbf{\# Types}  & \textbf{112} & \textbf{40}  \\ \hline
\end{tabular}
\renewcommand{\arraystretch}{1.0}
\end{table} 

\subsection{Prompt-based baseline methods for accessibility violation correction}
We compare \method \space to three other baselines, that use the following prompt engineering techniques to correct Web accessibility violations: contextual prompting \cite{DBLP:conf/petra/OthmanDJ23}, Re-Act prompting \cite{DBLP:journals/corr/abs-2401-16450}, and zero-shot prompting \cite{DBLP:conf/ccnc/DelnevoAM24}, using our sampled dataset and the dataset from Huang et al. \cite{DBLP:journals/corr/abs-2401-16450} (See section \ref{relatedWork}). 

\subsection{Implementation}
We implement the \method \emph{Detect} using \\ \texttt{Axe-Playwright-1.51.0} for syntax and layout accessibility violations and \texttt{GPT-4o} for semantic accessibility violations requiring multimodal reasoning capabilities to reason over Web page screenshot (See Section \ref{detect}).
For \method \emph{Correct}, we evaluate three LLMs on syntax and layout accessibility accessibility violations: \texttt{GPT-4-0125-preview}, \texttt{Mistral-7B-v0.1}, and \texttt{Qwen2.5-Coder}—a model optimized for coding tasks. These models vary in size, architecture, and training methods, enabling comparison across capabilities. 
To correct semantic accessibility violations requiring multimodal reasoning capabilities to reason over images (See Section \ref{correctt} and Table \ref{semanticExample}), we use: \texttt{GPT-4}, \texttt{Pixtral-12B}, and \texttt{Qwen-VL}.

\textbf{Handling Unreliable LLM Outputs:} During the manual inspection of LLM-generated responses, we identified several reliability issues in LLM-generated corrections. Some outputs were incomplete, hallucinated unrelated HTML, especially for long or multilingual pages (e.g., Chinese). In other cases, the model returned only textual advice without any HTML code. To ensure a fair and robust evaluation across all methods, we apply consistency checks to verify that LLM-generated corrections are valid and complete. For our method, we explicitly instruct the LLM to enclose the corrected code between string markers (e.g., \#\#\#START\#\#\# and \#\#\#END\#\#\#), enabling reliable extraction via regular expressions. For baselines that do not follow this format, we extract the first valid HTML snippet from the response heuristically, using pattern matching for common HTML tags.

For all LLM responses in our experiments, we manually verify that the extracted HTML contains a structurally valid and complete correction. If the response is missing required elements, contains malformed markup, or consists solely of textual advice, we flag it as not fixed. In these cases, the original violation score of the violation $v$ is used as the final output score to avoid inflating performance metrics.
\subsection{Evaluation Metrics}
We evaluate the effectiveness of \method{} in detecting and correcting Web accessibility violations across the three categories. The selection of evaluation metrics is informed by our proposed Web accessibility violation taxonomy, ensuring that each category is assessed using metrics appropriate to its specific characteristics. For each category, we employ category-specific metrics to assess whether the LLM-generated corrections resolve the detected accessibility violations, as measured by automated evaluation tools or manual validation.

\subsubsection{Detection Evaluation Metrics, } \label{detection_metrics}
We evaluate accessibility violation detection performance by measuring the detected violation count, the total number of Web accessibility violations identified across syntactic, layout, and semantic categories. 
\subsubsection{Syntactic and Layout Correction Metrics, } \label{snl}

we evaluate syntactic and layout accessibility compliance based on the reduction in violation scores. To calculate the average violation score of the entire dataset \(R\), we use Equation \ref{eq:dataset}, where \(i\) represents the index of each violation and \(n\) denotes the total number of entries in the dataset. 

\begin{equation}
  \label{eq:dataset}
  R = \frac{1}{n} \sum_{i=1}^{n} \textit{ViolationScore}(v_i)
\end{equation}
To calculate the percentage improvement \(I\) in violation score, we use Equation \ref{eq:two}, we compare the average violation scores before \(R_{\textit{initial}}\)  and after \(R_{\textit{fix}}\) applying a correction method.

\begin{equation}
\label{eq:two}
 I = 1 - \frac{R_{\textit{fix}}}{R_{\textit{initial}}}
\end{equation}

\subsubsection{Semantic Correction Metrics, } we calculate the similarity between human-generated and LLM-generated attributes using Sentence-BERT cosine similarity to assess the semantic quality of LLM-generated corrections. Sentence-BERT is well-suited for this task as it captures semantic meaning \cite{reimers2019sentence}, making it more robust than traditional word-overlap metrics such as BLEU \cite{post2018call} or ROUGE \cite{lin2004rouge}, which rely on exact n-gram matches. We then computed the average similarity score to evaluate the alignment between human and LLM-generated attributes. For instance, given an LLM-generated alt text attribute:
\textit{"A golden retriever playing with a ball in a grassy park."} We compare it to three human-generated variants:
\begin{itemize}
    \item \textit{"A dog fetching a ball in a green field."} (Similarity: 0.5986)
\item \textit{"A golden retriever running in the park with a toy."} (Similarity: 0.8364)
\item \textit{"A happy dog playing outdoors with a ball." }(Similarity: 0.6760)
\end{itemize}
The average similarity score across these responses is 0.7037, indicating a strong semantic alignment. 
\section{Experiments and Results}
\subsection{Detecting Accessibility Violations Experiments and Results (RQ1) }
To evaluate \textbf{RQ1—}, we applied \method \emph{Detect} to the HTML documents of the 16 URLs from the dataset by Huang et al.\cite{DBLP:journals/corr/abs-2401-16450}. Table \ref{tab:detection_coverage} compares the number of detected violations by \method{} \emph{Detect} against those originally reported in the dataset \cite{DBLP:journals/corr/abs-2401-16450},  \method{} \emph{Detect} outperforms the original dataset by detecting 104 semantic and more layout violations (15 vs. 8).

Although both methods use Axe-Playwright for detecting syntax and layout violations, \method{}\emph{Detect} reports fewer syntax (82 vs. 118) and more layout violations (15 vs. 8). To investigate this discrepancy, we refer to the counts reported in \cite{DBLP:journals/corr/abs-2401-16450}, which help explain the decrease in violations detected by automatic tools. Since the dataset from \cite{DBLP:journals/corr/abs-2401-16450} was collected in early 2024, some changes in reported Web accessibility violations are expected due to the dynamic nature of Web content. This interpretation is consistent with broader trends: according to the WebAIM Million study~\cite{Webaim2025}, Web accessibility has seen incremental improvements over the past year. For example, low color contrast violations decreased from 81\% in 2024 to 79.1\% in 2025.


\renewcommand{\arraystretch}{1.2}
\begin{table}[h]
\centering
\caption{Accessibility violation detection coverage on 16 pages from Huang et al.~\cite{DBLP:journals/corr/abs-2401-16450}.}
\label{tab:detection_coverage}
\small
\begin{tabular}{|l|c|c|}
\hline
\textbf{Violation Category} & \makecell{\textbf{Detected by} \\ \textbf{\cite{DBLP:journals/corr/abs-2401-16450}}} & \makecell{\textbf{Detected by} \\ \textbf{\method{} \emph{Detect}}} \\
\hline
Syntax    & 118 & 82  \\
Layout    & 8   & 15  \\
Semantic  & 0   & 104 \\
\hline
\textbf{Total} & \textbf{126} & \textbf{201} \\
\hline
\end{tabular}
\renewcommand{\arraystretch}{1.0}
\end{table}

\subsection{Syntactic and Layout Correction Experiments and Results (RQ2)}

To evaluate \textbf{RQ2—}, we conduct three analyses: (1) a comparison against three correction baselines, (2) a cross-LLM evaluation, and (3) an ablation study to assess the impact of our corrective re-prompting strategy.

\paragraph{Baselines Comparison.} We compare \method{} with three prompting-based baselines: contextual prompting~\cite{DBLP:conf/petra/OthmanDJ23}, ReAct prompting~\cite{DBLP:journals/corr/abs-2401-16450}, and zero-shot prompting~\cite{DBLP:conf/ccnc/DelnevoAM24}. As shown in Table~\ref{tab:syntax_layout_results}, \method{} consistently achieves the highest violation score decrease and number of corrected violations on both our dataset and the Huang et al. dataset~\cite{DBLP:journals/corr/abs-2401-16450}. With GPT-4, \method{} reduces violation scores by 0.84 (204 corrections) on our dataset, significantly outperforming ReAct and contextual prompting (0.50 and 0.46, respectively). On the Huang et al. dataset with GPT-4, \method{} reduces violation scores by 0.83 (141 corrections), significantly outperforming ReAct and contextual prompting (0.48 and 0.42, respectively).

\paragraph{Cross-LLM Comparison.} \method{} was tested with GPT-4, Qwen2.5, and Mistral-7B. GPT-4 consistently achieved the best performance across all correction tasks as shown in Table~\ref{tab:syntax_layout_results}.

\paragraph{Ablation Study.} To assess the impact of the corrective re-prompting strategy, we executed \method{} without the re-prompting phase using GPT-4. Performance dropped from 0.84 to 0.72 on our dataset, confirming the added value of this component. Even without it, \method{} outperforms the best-performing baseline, Re-Act prompting (0.50). This pattern holds consistently across all LLMs evaluated.

\paragraph{Qualitative Analysis.}  During the manual inspection of the results, we observed that all three baselines often provided incomplete solutions or adopted an "Occam's Razor" approach, where problematic elements were removed rather than properly corrected. This was frequently observed in long HTML snippets; when asked to correct an HTML snippet with sixteen elements, the output would only contain nine. In contrast, our method did not exhibit these issues. By asking the LLM to generate the code between string markers (\texttt{\#\#\#START\#\#\#} and \texttt{\#\#\#END\#\#\#}) in our prompts as shown in Table \ref{fig:mp}, we improved the model's ability to provide complete solutions. We also observed that baseline methods would occasionally resolve color contrast accessibility violations by changing both background and foreground colors to black and white. While this resolves the accessibility violation for users with diverse abilities, it distorts the visual design and layout for normal users. This behavior was not observed in our method's results, as our prompt template—shown in Table \ref{fig:mp}—was driven by our proposed taxonomy. The taxonomy entails that correcting layout accessibility violations shouldn't distort the visual design and layout for all users. Additionally, we attribute the notably poor performance of the zero-shot prompting baseline \cite{DBLP:conf/ccnc/DelnevoAM24} to the nature of its prompt design shown in Table \ref{tab:baseline_prompt_templates}. The prompt asked, \texttt{“Is the following HTML code accessible?”} without distinguishing between detection and correction tasks. As a result, we observed that the LLM often failed to recognize existing accessibility violations, and in cases where it did, it sometimes responded with only a confirmation that a violation was present, without generating the corrected code. This ambiguity in the prompt limited the effectiveness of the zero-shot approach across both datasets. Additionally, the lack of explicit reference to accessibility guidelines often leads to corrections that don't comply with recognized standards.



 \renewcommand{\arraystretch}{1.3}
\begin{table*}[t]
\centering
\caption{Comparison of violation score decrease and number of corrected accessibility violations for \textbf{syntax \& layout} violations on our dataset and the dataset from \cite{DBLP:journals/corr/abs-2401-16450}.}
\label{tab:syntax_layout_results}
\setlength{\tabcolsep}{4pt}
\resizebox{\textwidth}{!}{
\begin{tabular}{|p{5cm}|c|cc|cc|}
\hline
\multirow{3}{*}{\textbf{Method}} & 
\multirow{3}{*}{\textbf{Model}} & 
\multicolumn{2}{c|}{\textbf{Our Dataset (Size=250)}} & 
\multicolumn{2}{c|}{\textbf{Huang et al. Dataset \cite{DBLP:journals/corr/abs-2401-16450} (Size=171)}} \\
\cline{3-6}
& & 
\makecell{\textbf{Avg. Violation} \\ \textbf{Score Decrease}} & \makecell{\textbf{\# Corrected} \\ \textbf{Violations}} & 
\makecell{\textbf{Avg. Violation} \\ \textbf{Score Decrease}} & \makecell{\textbf{\# Corrected} \\ \textbf{Violations}} \\
\hline

Contextual prompting \cite{DBLP:conf/petra/OthmanDJ23} & GPT-4 & 0.46 & 123 & 0.42 & 86 \\
ReAct prompting \cite{DBLP:journals/corr/abs-2401-16450} & GPT-4 & 0.50 & 141 & 0.48 & 91 \\
Zero-shot prompting \cite{DBLP:conf/ccnc/DelnevoAM24} & GPT-4 & 0.19 & 43 & 0.12 & 19 \\
\textbf{\method} w/o reprompting (Ours) & GPT-4 & \underline{0.72} & \underline{184} & \underline{0.67} & \underline{119} \\
\textbf{\method (Ours)} & GPT-4 & \textbf{0.84} & \textbf{204} & \textbf{0.83} & \textbf{141} \\
\hline

Contextual prompting \cite{DBLP:conf/petra/OthmanDJ23} & Mistral-7B & 0.12 & 44 & 0.27 & 62 \\
ReAct prompting \cite{DBLP:journals/corr/abs-2401-16450} & Mistral-7B & 0.13 & 45 & 0.26 & 48 \\
Zero-shot prompting \cite{DBLP:conf/ccnc/DelnevoAM24} & Mistral-7B & 0.05 & 10 & 0.002 & 2 \\
\textbf{\method} w/o reprompting (Ours) & Mistral-7B & \underline{0.50} & \underline{162} & \underline{0.51} & \underline{110} \\
\textbf{\method \space (Ours)} & Mistral-7B & \textbf{0.82} & \textbf{200} & \textbf{0.76} & \textbf{127} \\
\hline

Contextual prompting \cite{DBLP:conf/petra/OthmanDJ23} & Qwen2.5 & 0.41 & 121 & 0.39 & 77 \\
ReAct prompting \cite{DBLP:journals/corr/abs-2401-16450} & Qwen2.5 & 0.44 & 130 & 0.37 & 71 \\
Zero-shot prompting \cite{DBLP:conf/ccnc/DelnevoAM24} & Qwen2.5 & 0.14 & 54 & 0.19 & 49 \\
\textbf{\method} w/o reprompting (Ours) & Qwen2.5 & \underline{0.49} & \underline{153} & \underline{0.52} & \underline{103} \\
\textbf{\method \space (Ours)} & Qwen2.5 & \textbf{0.74} & \textbf{183} & \textbf{0.75} & \textbf{126} \\
\hline

\end{tabular}
}
\end{table*}

\subsection{Semantic Correction Experiments and Results (RQ3)}

To evaluate \textbf{RQ3—}, we conducted two complementary evaluations: (1) human annotation to assess WCAG compliance and (2) a developer correction study to compare LLM corrections with those of human experts.

\paragraph{Human Annotation for WCAG Compliance.}
Two human annotators with five years of Web development and accessibility experience reviewed corrections for 55 semantic violations from our sampled dataset. The human annotators assigned a violation score of 0 if the correction fully resolved the accessibility violation according to WCAG 2.1; otherwise, the original violation score assigned during the detection of the semantic violation was retained. For example, if an image originally had the alt text “image” (violation score 5), and the LLM corrected it to a descriptive alt text, the violation was considered resolved, and the annotator assigned the correction a score of 0. If the correction is still vague (e.g., “image alt text”), the human annotator assigns a correction violation score of 5.  We then computed the average violation score decrease across all samples—i.e., how much the violation score was reduced after correction. As shown in Table~\ref{tab:semantic_results}, \method{} with GPT-4 achieved the highest average violation score decrease (0.96), resolving 53 out of 55 violations. This outperformed both ReAct prompting (0.87) and contextual prompting (0.82), confirming the effectiveness of our approach for semantic correction.

\paragraph{Qualitative Analysis.}
We also manually reviewed the semantic correction quality across GPT-4, Qwen-VL, and Pixtral, each showing distinct strengths. We found that explicitly instructing the model to consider the screenshot in the prompt—shown in Tables~\ref{fig:mp}, \ref{fig:mp_rePrompt},\ref{tab:baseline_prompt_templates}—was critical for eliciting image-grounded responses across all models. GPT-4 offered the most balanced performance, accurately integrating HTML structure and visual cues with minimal hallucinations. Qwen-VL showed strong image-based reasoning but frequently hallucinated additional HTML or introduced unrelated structures. Pixtral, by contrast, preserved original HTML faithfully and avoided unnecessary changes but often failed to ground its corrections in the provided image. For example, when prompted to generate alt text for the HTML5 logo, it produced generic attributes such as: (1) “an orange and white SVG graphic,” rather than a meaningful description of the image’s content—e.g., “the HTML5 logo.”

\paragraph{Comparison with Human Developer Corrections.}
To assess how closely \method{} aligns with human correction behavior, we conducted a human developer correction study. This human developer correction study builds on findings from prior work~\cite{fathallah2024empowering}, which demonstrated that LLMs are capable of generating semantic corrections. In particular, that study showed that GPT-based models could effectively address the \texttt{video-caption-not-descriptive} violation with corrections comparable in quality to those written by human annotators. Three full-stack developers independently corrected the same 55 semantic violations. Each developer received the HTML, violation metadata, and WCAG guidance—identical to what \method{} receives. We measured the similarity between LLM- and human-generated corrections using Sentence-BERT cosine similarity. As shown in Table~\ref{tab:semantic_similarity}, \method{} (GPT-4) achieved an average semantic similarity score of 0.77 when compared to human-generated corrections, indicating that the LLM-produced outputs closely matched the phrasing, structure, and meaning of human-written solutions.

While overall similarity scores indicate strong semantic alignment, certain categories—most notably \texttt{link-text-mismatch} and \texttt{form-label-mismatch}—had lower scores. For \\ \texttt{link-text-mismatch}, this is likely due to the LLM’s inability to access the target of hyperlinks, especially when the destination is a downloadable file (e.g., \texttt{<a href="rep123.pdf">Click here</a>}). Without knowing the link's content, the model cannot generate a descriptive label like \texttt{Download quarterly report}. In contrast, when the link points to another section of the same page (e.g., \texttt{\#contact}), the LLM performs better by using the surrounding context. In the case of \texttt{form-label-mismatch}, the LLM sometimes fails to infer connections between labels and input fields.

A full comparison of all methods across LLMs and violation types is provided in Tables \ref{tab:syntax_layout_results} and \ref{tab:semantic_results}, showing that \method \space consistently outperforms baselines on both datasets and across syntactic, layout, and semantic accessibility violations.
 
\renewcommand{\arraystretch}{1.3}

\begin{table*}[t]
\centering
\caption{Comparison of violation score decrease and number of corrected accessibility violations for \textbf{semantic} violations on our dataset (Size=55).}
\label{tab:semantic_results}

\begin{tabular}{|p{5cm}|c|c|c|}
\hline
\textbf{Method} & \textbf{Model} & \makecell{\textbf{Avg. Violation} \textbf{Score Decrease}} & \makecell{\textbf{\# Corrected} \textbf{Violations}} \\
\hline

Contextual prompting \cite{DBLP:conf/petra/OthmanDJ23} & GPT-4 & 0.82 & 44 \\
ReAct prompting \cite{DBLP:journals/corr/abs-2401-16450} & GPT-4 & 0.87 & 48 \\
Zero-shot prompting \cite{DBLP:conf/ccnc/DelnevoAM24} & GPT-4 & 0.33 & 18 \\
\textbf{\method} w/o reprompting (Ours) & GPT-4 & \underline{0.92} & \underline{51} \\
\textbf{\method (Ours)} & GPT-4 & \textbf{0.96} & \textbf{53} \\
\hline

Contextual prompting \cite{DBLP:conf/petra/OthmanDJ23} & Pixtral & 0.75 & 41 \\
ReAct prompting \cite{DBLP:journals/corr/abs-2401-16450} & Pixtral & 0.81 & 44 \\
Zero-shot prompting \cite{DBLP:conf/ccnc/DelnevoAM24} & Pixtral & 0.54 & 29 \\
\textbf{\method} w/o reprompting (Ours) & Pixtral & \underline{0.83} & \underline{46} \\
\textbf{\method (Ours)} & Pixtral & \textbf{0.92} & \textbf{51} \\
\hline

Contextual prompting \cite{DBLP:conf/petra/OthmanDJ23} & Qwen-VL & 0.60 & 32 \\
ReAct prompting \cite{DBLP:journals/corr/abs-2401-16450} & Qwen-VL & 0.37 & 20 \\
Zero-shot prompting \cite{DBLP:conf/ccnc/DelnevoAM24} & Qwen-VL & 0.18 & 8 \\
\textbf{\method} w/o reprompting (Ours) & Qwen-VL & \underline{0.69} & \underline{37} \\
\textbf{\method (Ours)} & Qwen-VL & \textbf{0.75} & \textbf{41} \\
\hline

\end{tabular}

\end{table*}

\renewcommand{\arraystretch}{1.2}
\begin{table}[h]
\centering
\caption{S-BERT similarity between \method{} (\texttt{GPT-4}) and human corrections by semantic category.}
\label{tab:semantic_similarity}
\small
\begin{tabular}{|p{3.5cm}|c|c|}
\hline
\textbf{Violation Type} & \textbf{\#} & \textbf{Avg. Sim.} \\ \hline
image-alt-not-descriptive     & 6  & 0.83  \\ \hline
lang-mismatch                 & 18 & 0.84  \\ \hline
link-text-mismatch            & 11 & 0.68  \\ \hline
form-label-mismatch           & 5  & 0.70  \\ \hline
ambiguous-heading             & 4  & 0.68  \\ \hline
page-title-not-descriptive    & 3  & 0.86  \\ \hline
button-label-mismatch         & 8  & 0.83  \\ \hline
\multicolumn{2}{|r|}{\textbf{Avg. Across 55 Violations}} & \textbf{0.77} \\ \hline
\end{tabular}
\renewcommand{\arraystretch}{1.0}
\end{table}

\section{Limitations}

We have shown that \method \space works conceptually and have evaluated it thoroughly. However, several non-trivial engineering efforts have not yet been integrated into our current implementation. These include the reconstruction of a fully corrected HTML document. While our correction module outputs individual corrected segments per violation, reintegrating these into the original document requires resolving potential conflicts, particularly when multiple overlapping corrections affect nested or related elements. For instance, corrections to a list and an image within a list item must be merged carefully to preserve the semantic structure. 

Another limitation is the reliability of the LLM-based semantic detector; in our observations, the LLM occasionally hallucinated accessibility violations or misidentified affected elements due to a lack of grounding in HTML attributes or structures. Long HTML documents pose challenges; they increase the prompt length, potentially overwhelming the model's context window and leading to incomplete reasoning or fabricated results. The LLM-based semantic detector relies on a static screenshot of the Web page, which doesn't capture dynamic content or alternate views such as drop-down menus, pop-ups, or language toggles.

While we manually verified the correctness of detected violations, we did not establish the complete set of ground-truth violations for each Web page. As a result, we do not report recall or precision metrics. This limits our ability to quantify undetected violations and fully assess detection completeness.

\begin{figure*}
  \centering
    \caption{Breakdown of top 10 uncorrected Web accessibility violations by AccessGuru across three LLMs (GPT-4, Mistral, and Qwen).Violations include: (a) \texttt{page-has-heading-one} – Ensure the page, or at least one level-one heading;
(b) \texttt{color-contrast} – Ensure sufficient contrast between text and background per WCAG 2 AA.
(c) \texttt{link-name} – Ensure links have discernible text;
(d) \texttt{duplicate-id} – Ensure every id attribute value is unique;
(e) \texttt{region} – Ensure all page content is contained by landmarks;
(f) \texttt{landmark-one-main} – Ensure the document has a main landmark;
(g) \texttt{landmark-unique} – Ensure landmarks are unique;
(h) \texttt{landmark-no-duplicate-main} – Ensure the document has at most one main landmark;
(i) \texttt{role-img-alt} – Ensure elements with role="img" have alternative text;
(j) \texttt{aria-tooltip-name} – Ensure every ARIA tooltip node has an accessible name.}
    \Description{Breakdown of top 10 uncorrected Web accessibility violations by AccessGuru across three LLMs (GPT-4, Mistral, and Qwen).}
  \includegraphics[width=16cm, alt={Breakdown of top 10 uncorrected Web accessibility violations by AccessGuru across three LLMs (GPT-4, Mistral, and Qwen).}]{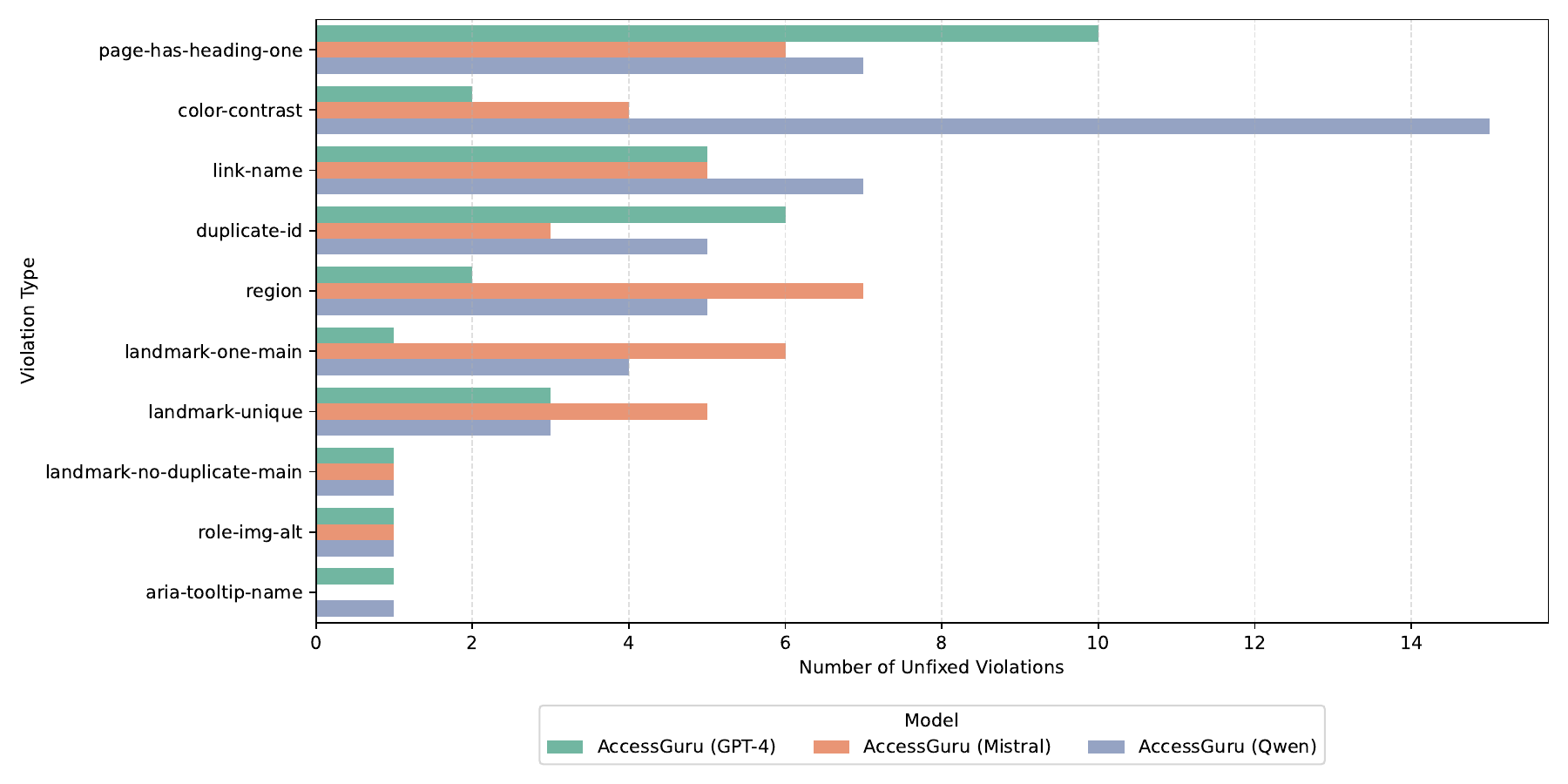}
  \label{breakdown}
\end{figure*}

Finally, correcting accessibility violations is exceptionally difficult, as it requires a nuanced understanding of user needs, Web design principles, and advanced reasoning capabilities to ensure effective correction \cite{baptista2016Web, craven2006Web}. To better understand which violations remained uncorrected, we analyzed the corrections produced by three \method{} models—GPT-4, Mistral, and Qwen. As shown in Figure \ref{breakdown}, certain violations, such as \texttt{page-has-heading-one}, \texttt{color-contrast}, and \texttt{link-name}, were consistently difficult to correct. 

Another limitation of \method\ is that it corrects \\ \texttt{color-contrast} accessibility violations by adjusting foreground and background values to meet WCAG thresholds only. However, it does not account for cases where color is used to convey meaning, such as red for errors or green for success. In such scenarios, even after contrast improvements, users with color vision deficiencies may still struggle to understand the intended message. Future work should investigate how visual cues like color can be supplemented with alternative representations (e.g., icons, text labels, or ARIA attributes) to ensure the semantic meaning is preserved for all users.

\section{Conclusion and Future Work}
Our proposed solution, \method, helps to improve Web accessibility by detecting and correcting accessibility violations in HTML documents. \method\ combines automated evaluation tools with prompting strategies for pre-trained LLMs to detect and generate compliant corrections. Central to \method\ is a taxonomy that defines violation types—\emph{Syntactic}, \emph{Layout}, and \emph{Semantic}—along with their required context and correction strategies, which guides both the detection and correction stages. 

\method\ \emph{Detect} is evaluated by comparing its accessibility violation coverage against existing accessibility detection datasets. \method\ \emph{Correct} is evaluated on two datasets using three different LLMs, with additional human studies to assess the semantic quality of the corrections compared to human-generated solutions. Our benchmark extends beyond automatic evaluation tools' conformance by addressing semantic accessibility violations. Future work should follow this direction by also moving beyond conformance checks and incorporating measures like task completion and usability, which automatic evaluation tools alone do not fully capture.

An important future direction is the reconstruction of fully corrected Web pages. While our system outputs individual corrected HTML snippets per violation, these must be re-integrated into the original document to form a coherent and fully accessible Web page. This reconstruction process involves merging overlapping or nested corrections while preserving the semantic structure, layout, and any unaffected content. We intend to address this engineering effort in future iterations of our system.

Additionally, future research should explore more robust grounding techniques, multi-view analysis of dynamic content—including the use of video to capture temporal changes—and strategies to handle long-context scenarios—such as hierarchical analysis or chunked evaluation—to improve the effectiveness of LLMs in detecting semantic accessibility violations. Finally, accessible Web pages available in the wild could be leveraged in future work to help LLMs retrieve compliant examples and apply similar patterns in corrections.

\section{Acknowledgements}
We acknowledge the support of the Stuttgart Research Focus Interchange Forum for Reflection on Intelligent Systems (IRIS).
\bibliographystyle{ACM-Reference-Format}
\bibliography{sample-base}


\begin{thebibliography}{60}


\ifx \showCODEN    \undefined \def \showCODEN     #1{\unskip}     \fi
\ifx \showISBNx    \undefined \def \showISBNx     #1{\unskip}     \fi
\ifx \showISBNxiii \undefined \def \showISBNxiii  #1{\unskip}     \fi
\ifx \showISSN     \undefined \def \showISSN      #1{\unskip}     \fi
\ifx \showLCCN     \undefined \def \showLCCN      #1{\unskip}     \fi
\ifx \shownote     \undefined \def \shownote      #1{#1}          \fi
\ifx \showarticletitle \undefined \def \showarticletitle #1{#1}   \fi
\ifx \showURL      \undefined \def \showURL       {\relax}        \fi
\providecommand\bibfield[2]{#2}
\providecommand\bibinfo[2]{#2}
\providecommand\natexlab[1]{#1}
\providecommand\showeprint[2][]{arXiv:#2}

\bibitem[{AChecker: IDI Accessibility}({[n.\,d.]})]%
        {achecker2024}
\bibfield{author}{\bibinfo{person}{{AChecker: IDI Accessibility}}.} \bibinfo{year}{[n.\,d.]}\natexlab{}.
\newblock \bibinfo{title}{AChecker: IDI Accessibility}.
\newblock \bibinfo{howpublished}{\url{http://www.atutor.ca/achecker/}}.
\newblock
\newblock
\shownote{Retrieved 25-06-2024}.


\bibitem[Acosta-Vargas et~al\mbox{.}(2020)]%
        {acosta2020dataset}
\bibfield{author}{\bibinfo{person}{Patricia Acosta-Vargas}, \bibinfo{person}{Mario Gonz{\'a}lez}, {and} \bibinfo{person}{Sergio Luj{\'a}n-Mora}.} \bibinfo{year}{2020}\natexlab{}.
\newblock \showarticletitle{Dataset for evaluating the accessibility of the websites of selected Latin American universities}.
\newblock \bibinfo{journal}{\emph{Data in brief}}  \bibinfo{volume}{28} (\bibinfo{year}{2020}), \bibinfo{pages}{105013}.
\newblock


\bibitem[Ahmad et~al\mbox{.}(2021)]%
        {DBLP:conf/naacl/AhmadCRC21}
\bibfield{author}{\bibinfo{person}{Wasi~Uddin Ahmad}, \bibinfo{person}{Saikat Chakraborty}, \bibinfo{person}{Baishakhi Ray}, {and} \bibinfo{person}{Kai{-}Wei Chang}.} \bibinfo{year}{2021}\natexlab{}.
\newblock \showarticletitle{Unified Pre-training for Program Understanding and Generation}. In \bibinfo{booktitle}{\emph{Proceedings of the 2021 Conference of the North American Chapter of the Association for Computational Linguistics: Human Language Technologies, {NAACL-HLT} 2021, Online, June 6-11, 2021}}, \bibfield{editor}{\bibinfo{person}{Kristina Toutanova}, \bibinfo{person}{Anna Rumshisky}, \bibinfo{person}{Luke Zettlemoyer}, \bibinfo{person}{Dilek Hakkani{-}T{\"{u}}r}, \bibinfo{person}{Iz~Beltagy}, \bibinfo{person}{Steven Bethard}, \bibinfo{person}{Ryan Cotterell}, \bibinfo{person}{Tanmoy Chakraborty}, {and} \bibinfo{person}{Yichao Zhou}} (Eds.). \bibinfo{publisher}{Association for Computational Linguistics}, \bibinfo{pages}{2655--2668}.
\newblock
\href{https://doi.org/10.18653/V1/2021.NAACL-MAIN.211}{doi:\nolinkurl{10.18653/V1/2021.NAACL-MAIN.211}}


\bibitem[Aljedaani et~al\mbox{.}(2024)]%
        {DBLP:conf/w4a/AljedaaniHAEF24}
\bibfield{author}{\bibinfo{person}{Wajdi Aljedaani}, \bibinfo{person}{Abdulrahman Habib}, \bibinfo{person}{Ahmed Aljohani}, \bibinfo{person}{Marcelo Eler}, {and} \bibinfo{person}{Yunhe Feng}.} \bibinfo{year}{2024}\natexlab{}.
\newblock \showarticletitle{Does ChatGPT Generate Accessible Code? Investigating Accessibility Challenges in LLM-Generated Source Code}. In \bibinfo{booktitle}{\emph{Proceedings of the 21st International Web for All Conference, {W4A} 2024, Singapore, May 13-14, 2024}}. \bibinfo{publisher}{{ACM}}, \bibinfo{pages}{165--176}.
\newblock
\href{https://doi.org/10.1145/3677846.3677854}{doi:\nolinkurl{10.1145/3677846.3677854}}


\bibitem[Almasoud and Mathkour(2019)]%
        {DBLP:conf/icisdm/AlmasoudM19}
\bibfield{author}{\bibinfo{person}{Suliman~K. Almasoud} {and} \bibinfo{person}{Hassan~I. Mathkour}.} \bibinfo{year}{2019}\natexlab{}.
\newblock \showarticletitle{Instant Adaptation Enrichment Technique to Improve Web Accessibility for Blind Users}. In \bibinfo{booktitle}{\emph{Proceedings of the 3rd International Conference on Information System and Data Mining, {ICISDM} 2019, Houston, TX, USA, April 6-8, 2019}}. \bibinfo{publisher}{{ACM}}, \bibinfo{pages}{159--164}.
\newblock
\href{https://doi.org/10.1145/3325917.3325931}{doi:\nolinkurl{10.1145/3325917.3325931}}


\bibitem[Alshayban et~al\mbox{.}(2020)]%
        {DBLP:conf/icse/Alshayban0M20}
\bibfield{author}{\bibinfo{person}{Abdulaziz Alshayban}, \bibinfo{person}{Iftekhar Ahmed}, {and} \bibinfo{person}{Sam Malek}.} \bibinfo{year}{2020}\natexlab{}.
\newblock \showarticletitle{Accessibility issues in Android apps: state of affairs, sentiments, and ways forward}. In \bibinfo{booktitle}{\emph{{ICSE} '20: 42nd International Conference on Software Engineering, Seoul, South Korea, 27 June - 19 July, 2020}}, \bibfield{editor}{\bibinfo{person}{Gregg Rothermel} {and} \bibinfo{person}{Doo{-}Hwan Bae}} (Eds.). \bibinfo{publisher}{{ACM}}, \bibinfo{pages}{1323--1334}.
\newblock
\href{https://doi.org/10.1145/3377811.3380392}{doi:\nolinkurl{10.1145/3377811.3380392}}


\bibitem[Austin et~al\mbox{.}(2021)]%
        {DBLP:journals/corr/abs-2108-07732}
\bibfield{author}{\bibinfo{person}{Jacob Austin}, \bibinfo{person}{Augustus Odena}, \bibinfo{person}{Maxwell~I. Nye}, \bibinfo{person}{Maarten Bosma}, \bibinfo{person}{Henryk Michalewski}, \bibinfo{person}{David Dohan}, \bibinfo{person}{Ellen Jiang}, \bibinfo{person}{Carrie~J. Cai}, \bibinfo{person}{Michael Terry}, \bibinfo{person}{Quoc~V. Le}, {and} \bibinfo{person}{Charles Sutton}.} \bibinfo{year}{2021}\natexlab{}.
\newblock \showarticletitle{Program Synthesis with Large Language Models}.
\newblock \bibinfo{journal}{\emph{CoRR}}  \bibinfo{volume}{abs/2108.07732} (\bibinfo{year}{2021}).
\newblock
\showeprint[arXiv]{2108.07732}
\urldef\tempurl%
\url{https://arxiv.org/abs/2108.07732}
\showURL{%
\tempurl}


\bibitem[Baptista et~al\mbox{.}(2016)]%
        {baptista2016Web}
\bibfield{author}{\bibinfo{person}{Ana Baptista}, \bibinfo{person}{Jos{\'e} Martins}, \bibinfo{person}{Ramiro Goncalves}, \bibinfo{person}{Frederico Branco}, {and} \bibinfo{person}{Tania Rocha}.} \bibinfo{year}{2016}\natexlab{}.
\newblock \showarticletitle{Web accessibility challenges and perspectives: A systematic literature review}. In \bibinfo{booktitle}{\emph{2016 11th Iberian Conference on Information Systems and Technologies (CISTI)}}. IEEE, \bibinfo{pages}{1--6}.
\newblock


\bibitem[Bi et~al\mbox{.}(2022)]%
        {DBLP:journals/tosem/BiXLGZF22}
\bibfield{author}{\bibinfo{person}{Tingting Bi}, \bibinfo{person}{Xin Xia}, \bibinfo{person}{David Lo}, \bibinfo{person}{John~C. Grundy}, \bibinfo{person}{Thomas Zimmermann}, {and} \bibinfo{person}{Denae Ford}.} \bibinfo{year}{2022}\natexlab{}.
\newblock \showarticletitle{Accessibility in Software Practice: {A} Practitioner's Perspective}.
\newblock \bibinfo{journal}{\emph{{ACM} Trans. Softw. Eng. Methodol.}} \bibinfo{volume}{31}, \bibinfo{number}{4} (\bibinfo{year}{2022}), \bibinfo{pages}{66:1--66:26}.
\newblock
\href{https://doi.org/10.1145/3503508}{doi:\nolinkurl{10.1145/3503508}}


\bibitem[Caldwell et~al\mbox{.}(2008)]%
        {caldwell2008Web}
\bibfield{author}{\bibinfo{person}{Ben Caldwell}, \bibinfo{person}{Michael Cooper}, \bibinfo{person}{Loretta~Guarino Reid}, \bibinfo{person}{Gregg Vanderheiden}, \bibinfo{person}{Wendy Chisholm}, \bibinfo{person}{John Slatin}, {and} \bibinfo{person}{Jason White}.} \bibinfo{year}{2008}\natexlab{}.
\newblock \showarticletitle{Web content accessibility guidelines (WCAG) 2.0}.
\newblock \bibinfo{journal}{\emph{WWW Consortium (W3C)}} \bibinfo{volume}{290}, \bibinfo{number}{1-34} (\bibinfo{year}{2008}), \bibinfo{pages}{5--12}.
\newblock


\bibitem[Cesarano et~al\mbox{.}(2007)]%
        {DBLP:conf/wse/CesaranoFT07}
\bibfield{author}{\bibinfo{person}{Carmine Cesarano}, \bibinfo{person}{Anna~Rita Fasolino}, {and} \bibinfo{person}{Porfirio Tramontana}.} \bibinfo{year}{2007}\natexlab{}.
\newblock \showarticletitle{Improving Usability of Web Pages for Blinds}. In \bibinfo{booktitle}{\emph{Proceedings of the 9th {IEEE} International Symposium on Web Systems Evolution, {WSE} 2009, 5-6 October 2007, Paris, France}}, \bibfield{editor}{\bibinfo{person}{Shihong Huang} {and} \bibinfo{person}{Massimiliano~Di Penta}} (Eds.). \bibinfo{publisher}{{IEEE} Computer Society}, \bibinfo{pages}{97--104}.
\newblock
\href{https://doi.org/10.1109/WSE.2007.4380250}{doi:\nolinkurl{10.1109/WSE.2007.4380250}}


\bibitem[Chen et~al\mbox{.}(2020)]%
        {DBLP:conf/icse/ChenCXXZ0W20}
\bibfield{author}{\bibinfo{person}{Jieshan Chen}, \bibinfo{person}{Chunyang Chen}, \bibinfo{person}{Zhenchang Xing}, \bibinfo{person}{Xiwei Xu}, \bibinfo{person}{Liming Zhu}, \bibinfo{person}{Guoqiang Li}, {and} \bibinfo{person}{Jinshui Wang}.} \bibinfo{year}{2020}\natexlab{}.
\newblock \showarticletitle{Unblind your apps: predicting natural-language labels for mobile {GUI} components by deep learning}. In \bibinfo{booktitle}{\emph{{ICSE} '20: 42nd International Conference on Software Engineering, Seoul, South Korea, 27 June - 19 July, 2020}}, \bibfield{editor}{\bibinfo{person}{Gregg Rothermel} {and} \bibinfo{person}{Doo{-}Hwan Bae}} (Eds.). \bibinfo{publisher}{{ACM}}, \bibinfo{pages}{322--334}.
\newblock
\href{https://doi.org/10.1145/3377811.3380327}{doi:\nolinkurl{10.1145/3377811.3380327}}


\bibitem[Chen et~al\mbox{.}(2021)]%
        {DBLP:journals/corr/abs-2107-03374}
\bibfield{author}{\bibinfo{person}{Mark Chen}, \bibinfo{person}{Jerry Tworek}, \bibinfo{person}{Heewoo Jun}, \bibinfo{person}{Qiming Yuan}, \bibinfo{person}{Henrique~Pond{\'{and}} de Oliveira~Pinto}, \bibinfo{person}{Jared Kaplan}, \bibinfo{person}{Harrison Edwards}, \bibinfo{person}{Yuri swear~here Nicholas~Joseph}, \bibinfo{person}{Greg Brockman}, \bibinfo{person}{Alex Ray}, \bibinfo{person}{Raul Puri}, \bibinfo{person}{Gretchen Krueger}, \bibinfo{person}{Michael Petrov}, \bibinfo{person}{Heidy Khlaaf}, \bibinfo{person}{Girish Sastry}, \bibinfo{person}{Pamela Mishkin}, \bibinfo{person}{Brooke Chan}, \bibinfo{person}{Scott Gray}, \bibinfo{person}{Nick Ryder}, \bibinfo{person}{Mikhail Pavlov}, \bibinfo{person}{Alethea Power}, \bibinfo{person}{Lukasz Kaiser}, \bibinfo{person}{Mohammad Bavarian}, \bibinfo{person}{Clemens Winter}, \bibinfo{person}{Philippe Tillet}, \bibinfo{person}{Felipe~Petroski Such}, \bibinfo{person}{Dave Cummings}, \bibinfo{person}{Matthias Plappert}, \bibinfo{person}{Fotios Chantzis},
  \bibinfo{person}{Elizabeth Barnes}, \bibinfo{person}{Ariel Herbert{-}Voss}, \bibinfo{person}{William~Hebgen Cast}, \bibinfo{person}{Alex Nichol}, \bibinfo{person}{Alex Paino}, \bibinfo{person}{Nikolas Tezak}, \bibinfo{person}{Jie Tang}, \bibinfo{person}{Igor Babuschkin}, \bibinfo{person}{Suchir Balaji}, \bibinfo{person}{Shantanu Jain}, \bibinfo{person}{William Saunders}, \bibinfo{person}{Christopher Hesse}, \bibinfo{person}{Andrew~N. Carr}, \bibinfo{person}{Jan Leike}, \bibinfo{person}{Joshua Achiam}, \bibinfo{person}{Vedant Misra}, \bibinfo{person}{Evan Morikawa}, \bibinfo{person}{Alec Radford}, \bibinfo{person}{Matthew Knight}, \bibinfo{person}{Miles Brundage}, \bibinfo{person}{Mira Murati}, \bibinfo{person}{Katie Mayer}, \bibinfo{person}{Peter Welinder}, \bibinfo{person}{Bob McGrew}, \bibinfo{person}{Dario Amodei}, \bibinfo{person}{Sam McCandlish}, \bibinfo{person}{Ilya Sutskever}, {and} \bibinfo{person}{Wojciech Zaremba}.} \bibinfo{year}{2021}\natexlab{}.
\newblock \showarticletitle{Evaluating Large Language Models Trained on Code}.
\newblock \bibinfo{journal}{\emph{CoRR}}  \bibinfo{volume}{abs/2107.03374} (\bibinfo{year}{2021}).
\newblock
\showeprint[arXiv]{2107.03374}
\urldef\tempurl%
\url{https://arxiv.org/abs/2107.03374}
\showURL{%
\tempurl}


\bibitem[Craven(2006)]%
        {craven2006Web}
\bibfield{author}{\bibinfo{person}{Jenny Craven}.} \bibinfo{year}{2006}\natexlab{}.
\newblock \showarticletitle{Web accessibility: A review of research and initiatives}.
\newblock  (\bibinfo{year}{2006}).
\newblock


\bibitem[Delnevo et~al\mbox{.}(2024)]%
        {DBLP:conf/ccnc/DelnevoAM24}
\bibfield{author}{\bibinfo{person}{Giovanni Delnevo}, \bibinfo{person}{Manuel Andruccioli}, {and} \bibinfo{person}{Silvia Mirri}.} \bibinfo{year}{2024}\natexlab{}.
\newblock \showarticletitle{On the Interaction with Large Language Models for Web Accessibility: Implications and Challenges}. In \bibinfo{booktitle}{\emph{21st {IEEE} Consumer Communications {\&} Networking Conference, {CCNC} 2024, Las Vegas, NV, USA, January 6-9, 2024}}. \bibinfo{publisher}{{IEEE}}, \bibinfo{pages}{1--6}.
\newblock
\href{https://doi.org/10.1109/CCNC51664.2024.10454680}{doi:\nolinkurl{10.1109/CCNC51664.2024.10454680}}


\bibitem[Doush and Qasem(2024)]%
        {iyad2024}
\bibfield{author}{\bibinfo{person}{Iyad~Abu Doush} {and} \bibinfo{person}{Reem Qasem}.} \bibinfo{year}{2024}\natexlab{}.
\newblock \showarticletitle{Evaluating AI-Generated Web Code for Accessibility Compliance: A Metric-Driven Approach}. In \bibinfo{booktitle}{\emph{Proceedings of Software Development and Technologies for Enhancing Accessibility and Fighting Info-exclusion (DSAI '24)}}. \bibinfo{publisher}{ACM}.
\newblock


\bibitem[Fathallah et~al\mbox{.}(2024a)]%
        {fathallah2024empowering}
\bibfield{author}{\bibinfo{person}{Nadeen Fathallah}, \bibinfo{person}{Monika Bhole}, {and} \bibinfo{person}{Steffen Staab}.} \bibinfo{year}{2024}\natexlab{a}.
\newblock \showarticletitle{Empowering the Deaf and Hard of Hearing Community: Enhancing Video Captions Using Large Language Models}. In \bibinfo{booktitle}{\emph{Proceedings of Software Development and Technologies for Enhancing Accessibility and Fighting Info-exclusion (DSAI '24)}}. \bibinfo{publisher}{ACM}, \bibinfo{address}{New York, USA}, \bibinfo{pages}{1--9}.
\newblock
\href{https://doi.org/10.48550/arXiv.2412.00342}{doi:\nolinkurl{10.48550/arXiv.2412.00342}}


\bibitem[Fathallah et~al\mbox{.}(2024b)]%
        {fathallah2024neon}
\bibfield{author}{\bibinfo{person}{Nadeen Fathallah}, \bibinfo{person}{Arunav Das}, \bibinfo{person}{Stefano De~Giorgis}, \bibinfo{person}{Andrea Poltronieri}, \bibinfo{person}{Peter Haase}, {and} \bibinfo{person}{Liubov Kovriguina}.} \bibinfo{year}{2024}\natexlab{b}.
\newblock \showarticletitle{NeOn-GPT: A Large Language Model-Powered Pipeline for Ontology Learning}. In \bibinfo{booktitle}{\emph{The Extended Semantic Web Conference}}.
\newblock


\bibitem[Ferati and Sulejmani(2016)]%
        {DBLP:conf/hci/FeratiS16}
\bibfield{author}{\bibinfo{person}{Mexhid Ferati} {and} \bibinfo{person}{Lirim Sulejmani}.} \bibinfo{year}{2016}\natexlab{}.
\newblock \showarticletitle{Automatic Adaptation Techniques to Increase the Web Accessibility for Blind Users}. In \bibinfo{booktitle}{\emph{{HCI} International 2016 - Posters' Extended Abstracts - 18th International Conference, {HCI} International 2016, Toronto, Canada, July 17-22, 2016, Proceedings, Part {II}}} \emph{(\bibinfo{series}{Communications in Computer and Information Science}, Vol.~\bibinfo{volume}{618})}, \bibfield{editor}{\bibinfo{person}{Constantine Stephanidis}} (Ed.). \bibinfo{publisher}{Springer}, \bibinfo{pages}{30--36}.
\newblock
\href{https://doi.org/10.1007/978-3-319-40542-1\_5}{doi:\nolinkurl{10.1007/978-3-319-40542-1\_5}}


\bibitem[Fleming and Lau(2014)]%
        {fleming2014measure}
\bibfield{author}{\bibinfo{person}{Stephen~M Fleming} {and} \bibinfo{person}{Hakwan~C Lau}.} \bibinfo{year}{2014}\natexlab{}.
\newblock \showarticletitle{How to measure metacognition}.
\newblock \bibinfo{journal}{\emph{Frontiers in human neuroscience}}  \bibinfo{volume}{8} (\bibinfo{year}{2014}), \bibinfo{pages}{443}.
\newblock


\bibitem[Frith(2012)]%
        {frith2012role}
\bibfield{author}{\bibinfo{person}{Chris~D Frith}.} \bibinfo{year}{2012}\natexlab{}.
\newblock \showarticletitle{The role of metacognition in human social interactions}.
\newblock \bibinfo{journal}{\emph{Philosophical Transactions of the Royal Society B: Biological Sciences}} \bibinfo{volume}{367}, \bibinfo{number}{1599} (\bibinfo{year}{2012}), \bibinfo{pages}{2213--2223}.
\newblock


\bibitem[Gu et~al\mbox{.}(2021)]%
        {DBLP:journals/corr/abs-2111-02643}
\bibfield{author}{\bibinfo{person}{Xiaodong Gu}, \bibinfo{person}{Kang~Min Yoo}, {and} \bibinfo{person}{Sang{-}Woo Lee}.} \bibinfo{year}{2021}\natexlab{}.
\newblock \showarticletitle{Response Generation with Context-Aware Prompt Learning}.
\newblock \bibinfo{journal}{\emph{CoRR}}  \bibinfo{volume}{abs/2111.02643} (\bibinfo{year}{2021}).
\newblock
\showeprint[arXiv]{2111.02643}
\urldef\tempurl%
\url{https://arxiv.org/abs/2111.02643}
\showURL{%
\tempurl}


\bibitem[Huang et~al\mbox{.}(2024)]%
        {DBLP:journals/corr/abs-2401-16450}
\bibfield{author}{\bibinfo{person}{Calista Huang}, \bibinfo{person}{Alyssa Ma}, \bibinfo{person}{Suchir Vyasamudri}, \bibinfo{person}{Eugenie Puype}, \bibinfo{person}{Sayem Kamal}, \bibinfo{person}{Juan~Belza Garcia}, \bibinfo{person}{Salar Cheema}, {and} \bibinfo{person}{Michael Lutz}.} \bibinfo{year}{2024}\natexlab{}.
\newblock \showarticletitle{{ACCESS:} Prompt Engineering for Automated Web Accessibility Violation Corrections}.
\newblock \bibinfo{journal}{\emph{CoRR}}  \bibinfo{volume}{abs/2401.16450} (\bibinfo{year}{2024}).
\newblock
\href{https://doi.org/10.48550/ARXIV.2401.16450}{doi:\nolinkurl{10.48550/ARXIV.2401.16450}}
\showeprint[arXiv]{2401.16450}


\bibitem[Kirkpatrick et~al\mbox{.}(2023)]%
        {kirkpatrick2023wcag}
\bibfield{author}{\bibinfo{person}{Andrew Kirkpatrick}, \bibinfo{person}{Joshue O'Connor}, \bibinfo{person}{Alastair Campbell}, {and} \bibinfo{person}{Michael Cooper}.} \bibinfo{year}{2023}\natexlab{}.
\newblock \bibinfo{booktitle}{\emph{Web Content Accessibility Guidelines (WCAG) 2.1}}.
\newblock \bibinfo{type}{Technical report}. \bibinfo{institution}{World Wide Web Consortium (W3C)}.
\newblock
\urldef\tempurl%
\url{https://www.w3.org/TR/2023/REC-WCAG21-20230921/}
\showURL{%
\tempurl}


\bibitem[Kodandaram et~al\mbox{.}(2024)]%
        {kodandaram2024enabling}
\bibfield{author}{\bibinfo{person}{Satwik~Ram Kodandaram}, \bibinfo{person}{Utku Uckun}, \bibinfo{person}{Xiaojun Bi}, \bibinfo{person}{IV Ramakrishnan}, {and} \bibinfo{person}{Vikas Ashok}.} \bibinfo{year}{2024}\natexlab{}.
\newblock \showarticletitle{Enabling Uniform Computer Interaction Experience for Blind Users through Large Language Models}. In \bibinfo{booktitle}{\emph{Proceedings of the 26th International ACM SIGACCESS Conference on Computers and Accessibility}}. \bibinfo{pages}{1--14}.
\newblock


\bibitem[Kong et~al\mbox{.}(2024)]%
        {roleplay}
\bibfield{author}{\bibinfo{person}{Aobo Kong}, \bibinfo{person}{Shiwan Zhao}, \bibinfo{person}{Hao Chen}, \bibinfo{person}{Qicheng Li}, \bibinfo{person}{Yong Qin}, \bibinfo{person}{Ruiqi Sun}, \bibinfo{person}{Xin Zhou}, \bibinfo{person}{Enzhi Wang}, {and} \bibinfo{person}{Xiaohang Dong}.} \bibinfo{year}{2024}\natexlab{}.
\newblock \showarticletitle{Better Zero-Shot Reasoning with Role-Play Prompting}. In \bibinfo{booktitle}{\emph{Proceedings of the 2024 Conference of the North American Chapter of the Association for Computational Linguistics: Human Language Technologies (Volume 1: Long Papers), {NAACL} 2024, Mexico City, Mexico, June 16-21, 2024}}, \bibfield{editor}{\bibinfo{person}{Kevin Duh}, \bibinfo{person}{Helena G{\'{o}}mez{-}Adorno}, {and} \bibinfo{person}{Steven Bethard}} (Eds.). \bibinfo{publisher}{Association for Computational Linguistics}, \bibinfo{pages}{4099--4113}.
\newblock
\href{https://doi.org/10.18653/V1/2024.NAACL-LONG.228}{doi:\nolinkurl{10.18653/V1/2024.NAACL-LONG.228}}


\bibitem[Li et~al\mbox{.}(2020)]%
        {DBLP:conf/emnlp/LiLHZLG20}
\bibfield{author}{\bibinfo{person}{Yang Li}, \bibinfo{person}{Gang Li}, \bibinfo{person}{Luheng He}, \bibinfo{person}{Jingjie Zheng}, \bibinfo{person}{Hong Li}, {and} \bibinfo{person}{Zhiwei Guan}.} \bibinfo{year}{2020}\natexlab{}.
\newblock \showarticletitle{Widget Captioning: Generating Natural Language Description for Mobile User Interface Elements}. In \bibinfo{booktitle}{\emph{Proceedings of the 2020 Conference on Empirical Methods in Natural Language Processing, {EMNLP} 2020, Online, November 16-20, 2020}}, \bibfield{editor}{\bibinfo{person}{Bonnie Webber}, \bibinfo{person}{Trevor Cohn}, \bibinfo{person}{Yulan He}, {and} \bibinfo{person}{Yang Liu}} (Eds.). \bibinfo{publisher}{Association for Computational Linguistics}, \bibinfo{pages}{5495--5510}.
\newblock
\href{https://doi.org/10.18653/V1/2020.EMNLP-MAIN.443}{doi:\nolinkurl{10.18653/V1/2020.EMNLP-MAIN.443}}


\bibitem[Lin(2004)]%
        {lin2004rouge}
\bibfield{author}{\bibinfo{person}{Chin-Yew Lin}.} \bibinfo{year}{2004}\natexlab{}.
\newblock \showarticletitle{Rouge: A package for automatic evaluation of summaries}. In \bibinfo{booktitle}{\emph{Text summarization branches out}}. \bibinfo{pages}{74--81}.
\newblock


\bibitem[Liu et~al\mbox{.}(2023)]%
        {DBLP:conf/nips/LiuXW023}
\bibfield{author}{\bibinfo{person}{Jiawei Liu}, \bibinfo{person}{Chunqiu~Steven Xia}, \bibinfo{person}{Yuyao Wang}, {and} \bibinfo{person}{Lingming Zhang}.} \bibinfo{year}{2023}\natexlab{}.
\newblock \showarticletitle{Is Your Code Generated by ChatGPT Really Correct? Rigorous Evaluation of Large Language Models for Code Generation}. In \bibinfo{booktitle}{\emph{Advances in Neural Information Processing Systems 36: Annual Conference on Neural Information Processing Systems 2023, NeurIPS 2023, New Orleans, LA, USA, December 10 - 16, 2023}}, \bibfield{editor}{\bibinfo{person}{Alice Oh}, \bibinfo{person}{Tristan Naumann}, \bibinfo{person}{Amir Globerson}, \bibinfo{person}{Kate Saenko}, \bibinfo{person}{Moritz Hardt}, {and} \bibinfo{person}{Sergey Levine}} (Eds.).
\newblock
\urldef\tempurl%
\url{http://papers.nips.cc/paper\_files/paper/2023/hash/43e9d647ccd3e4b7b5baab53f0368686-Abstract-Conference.html}
\showURL{%
\tempurl}


\bibitem[L{\'o}pez-Gil and Pereira(2024)]%
        {lopez2024turning}
\bibfield{author}{\bibinfo{person}{Juan-Miguel L{\'o}pez-Gil} {and} \bibinfo{person}{Juanan Pereira}.} \bibinfo{year}{2024}\natexlab{}.
\newblock \showarticletitle{Turning manual web accessibility success criteria into automatic: an LLM-based approach}.
\newblock \bibinfo{journal}{\emph{Universal Access in the Information Society}} (\bibinfo{year}{2024}), \bibinfo{pages}{1--16}.
\newblock


\bibitem[Mangiatordi and Lazzari(2018)]%
        {DBLP:conf/ccnc/MangiatordiL18}
\bibfield{author}{\bibinfo{person}{Andrea Mangiatordi} {and} \bibinfo{person}{Marco Lazzari}.} \bibinfo{year}{2018}\natexlab{}.
\newblock \showarticletitle{Combined use of artificial intelligence and crowdsourcing to provide alternative content for images on websites}. In \bibinfo{booktitle}{\emph{15th {IEEE} Annual Consumer Communications {\&} Networking Conference, {CCNC} 2018, Las Vegas, NV, USA, January 12-15, 2018}}. \bibinfo{publisher}{{IEEE}}, \bibinfo{pages}{1--6}.
\newblock
\href{https://doi.org/10.1109/CCNC.2018.8319312}{doi:\nolinkurl{10.1109/CCNC.2018.8319312}}


\bibitem[Mehendale and Walishetti(2024)]%
        {mehendale2024dexassist}
\bibfield{author}{\bibinfo{person}{Shridhar Mehendale} {and} \bibinfo{person}{Ankit Walishetti}.} \bibinfo{year}{2024}\natexlab{}.
\newblock \showarticletitle{DexAssist: A Voice-Enabled Dual-LLM Framework for Accessible Web Navigation}.
\newblock \bibinfo{journal}{\emph{arXiv preprint arXiv:2411.12214}} (\bibinfo{year}{2024}).
\newblock


\bibitem[{Microsoft}(2024)]%
        {PlaywrightAPI}
\bibfield{author}{\bibinfo{person}{{Microsoft}}.} \bibinfo{year}{2024}\natexlab{}.
\newblock \bibinfo{title}{Playwright API}.
\newblock \bibinfo{howpublished}{\url{https://playwright.dev}}.
\newblock
\newblock
\shownote{Accessed: 2024-06-30}.


\bibitem[Nam et~al\mbox{.}(2024)]%
        {DBLP:conf/icse/NamMHVM24}
\bibfield{author}{\bibinfo{person}{Daye Nam}, \bibinfo{person}{Andrew Macvean}, \bibinfo{person}{Vincent~J. Hellendoorn}, \bibinfo{person}{Bogdan Vasilescu}, {and} \bibinfo{person}{Brad~A. Myers}.} \bibinfo{year}{2024}\natexlab{}.
\newblock \showarticletitle{Using an {LLM} to Help With Code Understanding}. In \bibinfo{booktitle}{\emph{Proceedings of the 46th {IEEE/ACM} International Conference on Software Engineering, {ICSE} 2024, Lisbon, Portugal, April 14-20, 2024}}. \bibinfo{publisher}{{ACM}}, \bibinfo{pages}{97:1--97:13}.
\newblock
\href{https://doi.org/10.1145/3597503.3639187}{doi:\nolinkurl{10.1145/3597503.3639187}}


\bibitem[Njifenjou et~al\mbox{.}(2024)]%
        {DBLP:journals/corr/abs-2406-18460}
\bibfield{author}{\bibinfo{person}{Ahmed Njifenjou}, \bibinfo{person}{Virgile Sucal}, \bibinfo{person}{Bassam Jabaian}, {and} \bibinfo{person}{Fabrice Lef{\`{e}}vre}.} \bibinfo{year}{2024}\natexlab{}.
\newblock \showarticletitle{Role-Play Zero-Shot Prompting with Large Language Models for Open-Domain Human-Machine Conversation}.
\newblock \bibinfo{journal}{\emph{CoRR}}  \bibinfo{volume}{abs/2406.18460} (\bibinfo{year}{2024}).
\newblock
\href{https://doi.org/10.48550/ARXIV.2406.18460}{doi:\nolinkurl{10.48550/ARXIV.2406.18460}}
\showeprint[arXiv]{2406.18460}


\bibitem[OpenAI(2023)]%
        {DBLP:journals/corr/abs-2303-08774}
\bibfield{author}{\bibinfo{person}{OpenAI}.} \bibinfo{year}{2023}\natexlab{}.
\newblock \showarticletitle{{GPT-4} Technical Report}.
\newblock \bibinfo{journal}{\emph{CoRR}}  \bibinfo{volume}{abs/2303.08774} (\bibinfo{year}{2023}).
\newblock
\href{https://doi.org/10.48550/ARXIV.2303.08774}{doi:\nolinkurl{10.48550/ARXIV.2303.08774}}
\showeprint[arXiv]{2303.08774}


\bibitem[Othman et~al\mbox{.}(2023)]%
        {DBLP:conf/petra/OthmanDJ23}
\bibfield{author}{\bibinfo{person}{Achraf Othman}, \bibinfo{person}{Amira Dhouib}, {and} \bibinfo{person}{Aljazi Nasser~Al Jabor}.} \bibinfo{year}{2023}\natexlab{}.
\newblock \showarticletitle{Fostering websites accessibility: {A} case study on the use of the Large Language Models ChatGPT for automatic remediation}. In \bibinfo{booktitle}{\emph{Proceedings of the 16th International Conference on PErvasive Technologies Related to Assistive Environments, {PETRA} 2023, Corfu, Greece, July 5-7, 2023}}. \bibinfo{publisher}{{ACM}}, \bibinfo{pages}{707--713}.
\newblock
\href{https://doi.org/10.1145/3594806.3596542}{doi:\nolinkurl{10.1145/3594806.3596542}}


\bibitem[Pereira et~al\mbox{.}(2024)]%
        {DBLP:journals/taccess/PereiraGRGD24}
\bibfield{author}{\bibinfo{person}{Let{\'{\i}}cia~Seixas Pereira}, \bibinfo{person}{Jo{\~{a}}o Guerreiro}, \bibinfo{person}{Andr{\'{e}} Rodrigues}, \bibinfo{person}{Tiago~Jo{\~{a}}o Guerreiro}, {and} \bibinfo{person}{Carlos Duarte}.} \bibinfo{year}{2024}\natexlab{}.
\newblock \showarticletitle{From Automation to User Empowerment: Investigating the Role of a Semi-automatic Tool in Social Media Accessibility}.
\newblock \bibinfo{journal}{\emph{{ACM} Trans. Access. Comput.}} \bibinfo{volume}{17}, \bibinfo{number}{3} (\bibinfo{year}{2024}), \bibinfo{pages}{13:1--13:25}.
\newblock
\href{https://doi.org/10.1145/3647643}{doi:\nolinkurl{10.1145/3647643}}


\bibitem[Post(2018)]%
        {post2018call}
\bibfield{author}{\bibinfo{person}{Matt Post}.} \bibinfo{year}{2018}\natexlab{}.
\newblock \showarticletitle{A call for clarity in reporting BLEU scores}.
\newblock \bibinfo{journal}{\emph{arXiv preprint arXiv:1804.08771}} (\bibinfo{year}{2018}).
\newblock


\bibitem[Raman et~al\mbox{.}(2022)]%
        {DBLP:journals/corr/abs-2211-09935}
\bibfield{author}{\bibinfo{person}{Shreyas~Sundara Raman}, \bibinfo{person}{Vanya Cohen}, \bibinfo{person}{Eric Rosen}, \bibinfo{person}{Ifrah Idrees}, \bibinfo{person}{David Paulius}, {and} \bibinfo{person}{Stefanie Tellex}.} \bibinfo{year}{2022}\natexlab{}.
\newblock \showarticletitle{Planning with Large Language Models via Corrective Re-prompting}.
\newblock \bibinfo{journal}{\emph{CoRR}}  \bibinfo{volume}{abs/2211.09935} (\bibinfo{year}{2022}).
\newblock
\href{https://doi.org/10.48550/ARXIV.2211.09935}{doi:\nolinkurl{10.48550/ARXIV.2211.09935}}
\showeprint[arXiv]{2211.09935}


\bibitem[Rao et~al\mbox{.}(2022)]%
        {DBLP:conf/cvpr/RaoZ0TZH0L22}
\bibfield{author}{\bibinfo{person}{Yongming Rao}, \bibinfo{person}{Wenliang Zhao}, \bibinfo{person}{Guangyi Chen}, \bibinfo{person}{Yansong Tang}, \bibinfo{person}{Zheng Zhu}, \bibinfo{person}{Guan Huang}, \bibinfo{person}{Jie Zhou}, {and} \bibinfo{person}{Jiwen Lu}.} \bibinfo{year}{2022}\natexlab{}.
\newblock \showarticletitle{DenseCLIP: Language-Guided Dense Prediction with Context-Aware Prompting}. In \bibinfo{booktitle}{\emph{{IEEE/CVF} Conference on Computer Vision and Pattern Recognition, {CVPR} 2022, New Orleans, LA, USA, June 18-24, 2022}}. \bibinfo{publisher}{{IEEE}}, \bibinfo{pages}{18061--18070}.
\newblock
\href{https://doi.org/10.1109/CVPR52688.2022.01755}{doi:\nolinkurl{10.1109/CVPR52688.2022.01755}}


\bibitem[Reimers and Gurevych(2019)]%
        {reimers2019sentence}
\bibfield{author}{\bibinfo{person}{Nils Reimers} {and} \bibinfo{person}{Iryna Gurevych}.} \bibinfo{year}{2019}\natexlab{}.
\newblock \showarticletitle{Sentence-bert: Sentence embeddings using siamese bert-networks}.
\newblock \bibinfo{journal}{\emph{arXiv preprint arXiv:1908.10084}} (\bibinfo{year}{2019}).
\newblock


\bibitem[Rutter et~al\mbox{.}(2007)]%
        {rutter2007Web}
\bibfield{author}{\bibinfo{person}{Richard Rutter}, \bibinfo{person}{Patrick~H Lauke}, \bibinfo{person}{Cynthia Waddell}, \bibinfo{person}{Jim Thatcher}, \bibinfo{person}{Shawn~Lawton Henry}, \bibinfo{person}{Bruce Lawson}, \bibinfo{person}{Andrew Kirkpatrick}, \bibinfo{person}{Christian Heilmann}, \bibinfo{person}{Michael~R Burks}, \bibinfo{person}{Bob Regan}, {et~al\mbox{.}}} \bibinfo{year}{2007}\natexlab{}.
\newblock \bibinfo{booktitle}{\emph{Web accessibility: Web standards and regulatory compliance}}.
\newblock \bibinfo{publisher}{Apress}.
\newblock


\bibitem[Sahoo et~al\mbox{.}(2024)]%
        {DBLP:journals/corr/abs-2402-07927}
\bibfield{author}{\bibinfo{person}{Pranab Sahoo}, \bibinfo{person}{Ayush~Kumar Singh}, \bibinfo{person}{Sriparna Saha}, \bibinfo{person}{Vinija Jain}, \bibinfo{person}{Samrat Mondal}, {and} \bibinfo{person}{Aman Chadha}.} \bibinfo{year}{2024}\natexlab{}.
\newblock \showarticletitle{A Systematic Survey of Prompt Engineering in Large Language Models: Techniques and Applications}.
\newblock \bibinfo{journal}{\emph{CoRR}}  \bibinfo{volume}{abs/2402.07927} (\bibinfo{year}{2024}).
\newblock
\href{https://doi.org/10.48550/ARXIV.2402.07927}{doi:\nolinkurl{10.48550/ARXIV.2402.07927}}
\showeprint[arXiv]{2402.07927}


\bibitem[Salisbury et~al\mbox{.}(2017)]%
        {DBLP:conf/hcomp/SalisburyKM17}
\bibfield{author}{\bibinfo{person}{Elliot Salisbury}, \bibinfo{person}{Ece Kamar}, {and} \bibinfo{person}{Meredith~Ringel Morris}.} \bibinfo{year}{2017}\natexlab{}.
\newblock \showarticletitle{Toward Scalable Social Alt Text: Conversational Crowdsourcing as a Tool for Refining Vision-to-Language Technology for the Blind}. In \bibinfo{booktitle}{\emph{Proceedings of the Fifth {AAAI} Conference on Human Computation and Crowdsourcing, {HCOMP} 2017, 23-26 October 2017, Qu{\'{e}}bec City, Qu{\'{e}}bec, Canada}}, \bibfield{editor}{\bibinfo{person}{Steven Dow} {and} \bibinfo{person}{Adam~Tauman Kalai}} (Eds.). \bibinfo{publisher}{{AAAI} Press}, \bibinfo{pages}{147--156}.
\newblock
\href{https://doi.org/10.1609/HCOMP.V5I1.13301}{doi:\nolinkurl{10.1609/HCOMP.V5I1.13301}}


\bibitem[Shanahan et~al\mbox{.}(2023)]%
        {DBLP:journals/nature/ShanahanMR23}
\bibfield{author}{\bibinfo{person}{Murray Shanahan}, \bibinfo{person}{Kyle McDonell}, {and} \bibinfo{person}{Laria Reynolds}.} \bibinfo{year}{2023}\natexlab{}.
\newblock \showarticletitle{Role play with large language models}.
\newblock \bibinfo{journal}{\emph{Nat.}} \bibinfo{volume}{623}, \bibinfo{number}{7987} (\bibinfo{year}{2023}), \bibinfo{pages}{493--498}.
\newblock
\href{https://doi.org/10.1038/S41586-023-06647-8}{doi:\nolinkurl{10.1038/S41586-023-06647-8}}


\bibitem[{\v{S}}mite et~al\mbox{.}(2014)]%
        {vsmite2014empirically}
\bibfield{author}{\bibinfo{person}{Darja {\v{S}}mite}, \bibinfo{person}{Claes Wohlin}, \bibinfo{person}{Zane Galvi{\c{n}}a}, {and} \bibinfo{person}{Rafael Prikladnicki}.} \bibinfo{year}{2014}\natexlab{}.
\newblock \showarticletitle{An empirically based terminology and taxonomy for global software engineering}.
\newblock \bibinfo{journal}{\emph{Empirical Software Engineering}}  \bibinfo{volume}{19} (\bibinfo{year}{2014}), \bibinfo{pages}{105--153}.
\newblock


\bibitem[Tuan et~al\mbox{.}(2012)]%
        {tuan2012checking}
\bibfield{author}{\bibinfo{person}{Dat~Trinh Tuan}, \bibinfo{person}{Van-Hung Phan}, {et~al\mbox{.}}} \bibinfo{year}{2012}\natexlab{}.
\newblock \showarticletitle{Checking and correcting the source code of web pages for accessibility}. In \bibinfo{booktitle}{\emph{2012 IEEE RIVF International Conference on Computing \& Communication Technologies, Research, Innovation, and Vision for the Future}}. IEEE, \bibinfo{pages}{1--4}.
\newblock


\bibitem[Usman et~al\mbox{.}(2017)]%
        {usman2017taxonomies}
\bibfield{author}{\bibinfo{person}{Muhammad Usman}, \bibinfo{person}{Ricardo Britto}, \bibinfo{person}{J{\"u}rgen B{\"o}rstler}, {and} \bibinfo{person}{Emilia Mendes}.} \bibinfo{year}{2017}\natexlab{}.
\newblock \showarticletitle{Taxonomies in software engineering: A systematic mapping study and a revised taxonomy development method}.
\newblock \bibinfo{journal}{\emph{Information and Software Technology}}  \bibinfo{volume}{85} (\bibinfo{year}{2017}), \bibinfo{pages}{43--59}.
\newblock


\bibitem[Vaithilingam et~al\mbox{.}(2022)]%
        {DBLP:conf/chi/Vaithilingam0G22}
\bibfield{author}{\bibinfo{person}{Priyan Vaithilingam}, \bibinfo{person}{Tianyi Zhang}, {and} \bibinfo{person}{Elena~L. Glassman}.} \bibinfo{year}{2022}\natexlab{}.
\newblock \showarticletitle{Expectation vs. Experience: Evaluating the Usability of Code Generation Tools Powered by Large Language Models}. In \bibinfo{booktitle}{\emph{{CHI} '22: {CHI} Conference on Human Factors in Computing Systems, New Orleans, LA, USA, 29 April 2022 - 5 May 2022, Extended Abstracts}}, \bibfield{editor}{\bibinfo{person}{Simone D.~J. Barbosa}, \bibinfo{person}{Cliff Lampe}, \bibinfo{person}{Caroline Appert}, {and} \bibinfo{person}{David~A. Shamma}} (Eds.). \bibinfo{publisher}{{ACM}}, \bibinfo{pages}{332:1--332:7}.
\newblock
\href{https://doi.org/10.1145/3491101.3519665}{doi:\nolinkurl{10.1145/3491101.3519665}}


\bibitem[Vinyals et~al\mbox{.}(2015)]%
        {DBLP:conf/cvpr/VinyalsTBE15}
\bibfield{author}{\bibinfo{person}{Oriol Vinyals}, \bibinfo{person}{Alexander Toshev}, \bibinfo{person}{Samy Bengio}, {and} \bibinfo{person}{Dumitru Erhan}.} \bibinfo{year}{2015}\natexlab{}.
\newblock \showarticletitle{Show and tell: {A} neural image caption generator}. In \bibinfo{booktitle}{\emph{{IEEE} Conference on Computer Vision and Pattern Recognition, {CVPR} 2015, Boston, MA, USA, June 7-12, 2015}}. \bibinfo{publisher}{{IEEE} Computer Society}, \bibinfo{pages}{3156--3164}.
\newblock
\href{https://doi.org/10.1109/CVPR.2015.7298935}{doi:\nolinkurl{10.1109/CVPR.2015.7298935}}


\bibitem[Wang et~al\mbox{.}(2024)]%
        {DBLP:conf/acl/WangPQLZWGGN00024}
\bibfield{author}{\bibinfo{person}{Noah Wang}, \bibinfo{person}{Z. y. Peng}, \bibinfo{person}{Haoran Que}, \bibinfo{person}{Jiaheng Liu}, \bibinfo{person}{Wangchunshu Zhou}, \bibinfo{person}{Yuhan Wu}, \bibinfo{person}{Hongcheng Guo}, \bibinfo{person}{Ruitong Gan}, \bibinfo{person}{Zehao Ni}, \bibinfo{person}{Jian Yang}, \bibinfo{person}{Man Zhang}, \bibinfo{person}{Zhaoxiang Zhang}, \bibinfo{person}{Wanli Ouyang}, \bibinfo{person}{Ke Xu}, \bibinfo{person}{Wenhao Huang}, \bibinfo{person}{Jie Fu}, {and} \bibinfo{person}{Junran Peng}.} \bibinfo{year}{2024}\natexlab{}.
\newblock \showarticletitle{RoleLLM: Benchmarking, Eliciting, and Enhancing Role-Playing Abilities of Large Language Models}. In \bibinfo{booktitle}{\emph{Findings of the Association for Computational Linguistics, {ACL} 2024, Bangkok, Thailand and virtual meeting, August 11-16, 2024}}, \bibfield{editor}{\bibinfo{person}{Lun{-}Wei Ku}, \bibinfo{person}{Andre Martins}, {and} \bibinfo{person}{Vivek Srikumar}} (Eds.). \bibinfo{publisher}{Association for Computational Linguistics}, \bibinfo{pages}{14743--14777}.
\newblock
\href{https://doi.org/10.18653/V1/2024.FINDINGS-ACL.878}{doi:\nolinkurl{10.18653/V1/2024.FINDINGS-ACL.878}}


\bibitem[Wang and Zhao(2023)]%
        {DBLP:journals/corr/abs-2308-05342}
\bibfield{author}{\bibinfo{person}{Yuqing Wang} {and} \bibinfo{person}{Yun Zhao}.} \bibinfo{year}{2023}\natexlab{}.
\newblock \showarticletitle{Metacognitive Prompting Improves Understanding in Large Language Models}.
\newblock \bibinfo{journal}{\emph{CoRR}}  \bibinfo{volume}{abs/2308.05342} (\bibinfo{year}{2023}).
\newblock
\href{https://doi.org/10.48550/ARXIV.2308.05342}{doi:\nolinkurl{10.48550/ARXIV.2308.05342}}
\showeprint[arXiv]{2308.05342}


\bibitem[{WebAIM}(2025)]%
        {Webaim2025wave}
\bibfield{author}{\bibinfo{person}{{WebAIM}}.} \bibinfo{year}{2025}\natexlab{}.
\newblock \bibinfo{title}{{WAVE}}.
\newblock \bibinfo{howpublished}{\url{https://wave.webaim.org/api/}}.
\newblock
\newblock
\shownote{Retrieved 2025-01-11}.


\bibitem[WebAIM(2025)]%
        {Webaim2025}
\bibfield{author}{\bibinfo{person}{WebAIM}.} \bibinfo{year}{2025}\natexlab{}.
\newblock \bibinfo{booktitle}{\emph{The WebAIM Million - An Annual Accessibility Analysis of the Top 1,000,000 Home Pages}}.
\newblock \bibinfo{type}{Technical Report}. \bibinfo{institution}{WebAIM.org}.
\newblock
\urldef\tempurl%
\url{https://webaim.org/projects/million/}
\showURL{%
\tempurl}


\bibitem[White et~al\mbox{.}(2023)]%
        {DBLP:journals/corr/abs-2302-11382}
\bibfield{author}{\bibinfo{person}{Jules White}, \bibinfo{person}{Quchen Fu}, \bibinfo{person}{Sam Hays}, \bibinfo{person}{Michael Sandborn}, \bibinfo{person}{Carlos Olea}, \bibinfo{person}{Henry Gilbert}, \bibinfo{person}{Ashraf Elnashar}, \bibinfo{person}{Jesse Spencer{-}Smith}, {and} \bibinfo{person}{Douglas~C. Schmidt}.} \bibinfo{year}{2023}\natexlab{}.
\newblock \showarticletitle{A Prompt Pattern Catalog to Enhance Prompt Engineering with ChatGPT}.
\newblock \bibinfo{journal}{\emph{CoRR}}  \bibinfo{volume}{abs/2302.11382} (\bibinfo{year}{2023}).
\newblock
\href{https://doi.org/10.48550/ARXIV.2302.11382}{doi:\nolinkurl{10.48550/ARXIV.2302.11382}}
\showeprint[arXiv]{2302.11382}


\bibitem[Wu et~al\mbox{.}(2017)]%
        {DBLP:conf/cscw/WuWFS17}
\bibfield{author}{\bibinfo{person}{Shaomei Wu}, \bibinfo{person}{Jeffrey Wieland}, \bibinfo{person}{Omid Farivar}, {and} \bibinfo{person}{Julie Schiller}.} \bibinfo{year}{2017}\natexlab{}.
\newblock \showarticletitle{Automatic Alt-text: Computer-generated Image Descriptions for Blind Users on a Social Network Service}. In \bibinfo{booktitle}{\emph{Proceedings of the 2017 {ACM} Conference on Computer Supported Cooperative Work and Social Computing, {CSCW} 2017, Portland, OR, USA, February 25 - March 1, 2017}}, \bibfield{editor}{\bibinfo{person}{Charlotte~P. Lee}, \bibinfo{person}{Steven~E. Poltrock}, \bibinfo{person}{Louise Barkhuus}, \bibinfo{person}{Marcos Borges}, {and} \bibinfo{person}{Wendy~A. Kellogg}} (Eds.). \bibinfo{publisher}{{ACM}}, \bibinfo{pages}{1180--1192}.
\newblock
\href{https://doi.org/10.1145/2998181.2998364}{doi:\nolinkurl{10.1145/2998181.2998364}}


\bibitem[Yesilada et~al\mbox{.}(2012)]%
        {DBLP:conf/w4a/YesiladaBVH12}
\bibfield{author}{\bibinfo{person}{Yeliz Yesilada}, \bibinfo{person}{Giorgio Brajnik}, \bibinfo{person}{Markel Vigo}, {and} \bibinfo{person}{Simon Harper}.} \bibinfo{year}{2012}\natexlab{}.
\newblock \showarticletitle{Understanding web accessibility and its drivers}. In \bibinfo{booktitle}{\emph{International Cross-Disciplinary Conference on Web Accessibility, {W4A} '12, Lyon, France, April 16-17, 2012}}, \bibfield{editor}{\bibinfo{person}{Markel Vigo}, \bibinfo{person}{Julio Abascal}, \bibinfo{person}{Rui Lopes}, {and} \bibinfo{person}{Paola Salomoni}} (Eds.). \bibinfo{publisher}{{ACM}}, \bibinfo{pages}{19}.
\newblock
\href{https://doi.org/10.1145/2207016.2207027}{doi:\nolinkurl{10.1145/2207016.2207027}}


\bibitem[Yesilada and Harper(2019)]%
        {DBLP:series/hci/YesiladaH19}
\bibfield{editor}{\bibinfo{person}{Yeliz Yesilada} {and} \bibinfo{person}{Simon Harper}} (Eds.). \bibinfo{year}{2019}\natexlab{}.
\newblock \bibinfo{booktitle}{\emph{Web Accessibility - {A} Foundation for Research, Second Edition}}.
\newblock \bibinfo{publisher}{Springer}.
\newblock
\showISBNx{978-1-4471-7439-4}
\href{https://doi.org/10.1007/978-1-4471-7440-0}{doi:\nolinkurl{10.1007/978-1-4471-7440-0}}


\bibitem[Zhou et~al\mbox{.}(2024)]%
        {DBLP:conf/www/0002LJND24}
\bibfield{author}{\bibinfo{person}{Yujia Zhou}, \bibinfo{person}{Zheng Liu}, \bibinfo{person}{Jiajie Jin}, \bibinfo{person}{Jian{-}Yun Nie}, {and} \bibinfo{person}{Zhicheng Dou}.} \bibinfo{year}{2024}\natexlab{}.
\newblock \showarticletitle{Metacognitive Retrieval-Augmented Large Language Models}. In \bibinfo{booktitle}{\emph{Proceedings of the {ACM} on Web Conference 2024, {WWW} 2024, Singapore, May 13-17, 2024}}, \bibfield{editor}{\bibinfo{person}{Tat{-}Seng Chua}, \bibinfo{person}{Chong{-}Wah Ngo}, \bibinfo{person}{Ravi Kumar}, \bibinfo{person}{Hady~W. Lauw}, {and} \bibinfo{person}{Roy~Ka{-}Wei Lee}} (Eds.). \bibinfo{publisher}{{ACM}}, \bibinfo{pages}{1453--1463}.
\newblock
\href{https://doi.org/10.1145/3589334.3645481}{doi:\nolinkurl{10.1145/3589334.3645481}}


\end{thebibliography}
\appendix
\section{Appendix} \label{appendix:prompt_templates}
This appendix complements our methodology by providing prompting templates and a subset of our proposed taxonomy for categorizing Web accessibility violations.

\renewcommand{\arraystretch}{1.2}  
\begin{table}[h]
\centering
\caption{Prompt template used in the LLM-based semantic detector within the \method{} \emph{Detect} module. The prompt guides the LLM to \textbf{detect semantic Web accessibility violations}.}
\label{tab:semantic_detection_prompt}
\small
\rowcolors{2}{white}{lightgray}
\resizebox{0.98\columnwidth}{!}{%
\begin{tabular}{p{1.3cm}p{6.2cm}}
\toprule
\textbf{Type} & \textbf{Prompt Template} \\
\midrule
Fixed & You are a Web accessibility expert. Your task is to detect semantic accessibility violations in the given HTML Web page.
These accessibility violations are often not detectable by standard automated tools and require interpretation of the content’s
meaning and user context \\
Fixed & A semantic violation occurs when: \newline
\hspace*{1em}- Attributes like alt text, language, or link/button labels are present but do not provide meaningful information.\newline
\hspace*{1em}- Visual or multimedia content is not described in a way that conveys its purpose to users with disabilities. \\
Fixed & Use the following context in your analysis: \newline
\hspace*{1em}- Domain: \\
Dynamic & \hspace*{1.5em}\texttt{\{Insert Web page Domain\}} \\
Fixed & \hspace*{1em}- URL: \\
Dynamic & \hspace*{1.5em}\texttt{\{Insert Web page URL\}} \\
Fixed & You are provided with: \newline
\hspace*{1em}- The HTML code of the Web page to analyze. \newline
\hspace*{1em}- The full semantic accessibility violation taxonomy. \newline
\texttt{[Semantic Accessibility Violation Taxonomy]} \newline
\hspace*{1em}- A screenshot of the rendered view of the Web page. \\
Dynamic & \hspace*{1.5em}\texttt{\{Insert HTML here\}} \\
Dynamic & \hspace*{1.5em}\texttt{\{Insert Web page Screenshot\}} \\
Fixed & Now, review the HTML and supplementary data. List all semantic accessibility violations you detect, and for each: \newline
1.  Identify the affected HTML element. Enclose the exact HTML snippet using the markers \texttt{[START]} and \texttt{[END]}. \newline
2. Specify the violation name. \\
\bottomrule
\end{tabular}
}
\renewcommand{\arraystretch}{1.0}
\end{table}

\renewcommand{\arraystretch}{1.2}
\begin{table}[h]
\centering
\caption{\textbf{Role-play persona} used in the \method{} \emph{Correct} module to guide the LLM in generating WCAG-compliant HTML corrections.}
\label{persona}
\small
\begin{tabular}{p{7.5cm}}
\toprule
\textbf{Persona} \\
\midrule
You are a Web accessibility expert with strong HTML skills and a deep commitment to fixing accessibility violations. You analyze Web pages, identify issues, and provide corrected HTML that meets WCAG 2.1 standards. \\
You resolve problems like missing alt text, poor heading structure, non-semantic elements, inaccessible forms, and color contrast issues. \\
You transform flawed code into compliant, clean HTML that works with assistive technologies and is fully keyboard-navigable and screen-reader-friendly. \\
Your mission is to deliver expert, immediately usable HTML fixes to make the Web inclusive for all users. \\
\bottomrule
\end{tabular}
\renewcommand{\arraystretch}{1.0}
\end{table}

\begin{table*}
\caption{\textbf{Initial prompt template} used in the \method \emph{Correct}  module. The prompt integrates role-play, contextual, and metacognitive prompting strategies and is structured around five metacognitive stages: comprehension clarification, preliminary judgment, critical evaluation, decision confirmation, and confidence assessment. The prompt is structured into fixed and dynamic components. The fixed components remain constant across all samples, while dynamic fields are populated based on the violation instance. (For semantic violations involving visual content, an additional instruction guides the LLM to reason over Webpage screenshots—see purple-highlighted rows.) }
\label{fig:mp}
\renewcommand{\arraystretch}{1.5}
\rowcolors{2}{white}{lightgray}
\centering
\resizebox{\textwidth}{!}{
\begin{tabular}{p{2.3cm}p{1cm}p{14cm}}
\toprule
\textbf{Metacognitive Stage} & \textbf{Type} & \textbf{Prompt Template} \\
\midrule
\multirow{2}{*}{\makecell{Comprehension \\  Clarification} }
& Fixed & \texttt{\{Role-play persona\}} + Clarify your understanding of the following Web accessibility violation: \textcolor{lightgray}{\texttt{(Role-play persona)} + Clarify your understanding of the following Web accessibility } \\
& Dynamic & \texttt{\{Category\}} \newline
\texttt{\{Category description\}} \newline
     \texttt{\{Violation name\}} \newline
      \texttt{\{Violation description\}} \newline
     \texttt{\{URL\}} \newline
     \texttt{\{HTMLElement\}}      \newline 
     \texttt{\{Impact\}} 
     \\ 
     & Fixed & Impact is a rating determined by the severity of the violation, indicating the extent to which it hinders user interaction with the Web content. The scale is [cosmetic, minor, moderate, serious, critical]\\ \rowcolor{blue!10}
     & Fixed & Prioritize the attached screenshot of the Web page (which visually shows the UI element with a possible Web accessibility violation).
  Your tasks:
  \begin{enumerate}
    \item Interpret the visual content of the attached image.
    \item Identify the UI element (e.g., a button or icon) shown in the image.
    \item  Determine whether the element is accessible (i.e., if an image element has a meaningful alt text)
    \item Compare your findings with the corresponding HTML provided and highlight any mismatches.
    \item Suggest an accessibility-compliant fix if there's a violation.
  \end{enumerate} \\ \rowcolor{blue!10}
  & Dynamic & \texttt{\{Web page screenshot\}} \\
     \hline
\multirow{2}{*}{\makecell{Preliminary \\  Judgment}} 
& Fixed &   Based on your understanding, provide a preliminary correction for the Web accessibility violation based on the following WCAG guideline(s): \textcolor{white}{violation based on the following WCAG guideline(s):}\\ 
& Dynamic & \texttt{\{Relevant WCAG guideline\}} \\ 
&Fixed &  Make sure your generated code corrects the Web accessibility violation without introducing new accessibility violations. Ensure you generate the complete corrected code, not just a snippet. \\ \hline

\multirow{2}{*}{\makecell{Critical \\ Evaluation} }
& Fixed &     Critically assess your preliminary correction, make sure to correct the initial Web accessibility violation without introducing new Web accessibility violations. Only make corrections if the previous answer is incorrect. Make sure your generated code corrects the Web accessibility violation without introducing new accessibility violations.\\ \hline

\multirow{2}{*}{\makecell{Decision \\ Confirmation}} 
& Fixed &  Confirm your final decision on whether the correction is accurate or not and provide the reasoning for your decision. Only suggest further corrections if the initial response contains errors. Make sure your generated code corrects the Web accessibility violation without introducing new accessibility violations. Enclose your corrected HTML code to replace the initial code with accessibility violations between these two marker strings: "\#\#\#START\#\#\#" as the first line and "\#\#\#END\#\#\#" as the last line.
  \\ \hline

\multirow{2}{*}{\makecell{Confidence \\ Assessment} }
& Fixed & Evaluate your confidence (0-100\%) in your correction, enclose your confidence score between these two marker strings: "\#\#\#START1\#\#\#" as first line and "\#\#\#END1\#\#\#" as last line. Provide an explanation for this confidence level; enclose your explanation between these two marker strings: "\#\#\#START2\#\#\#" as the first line and "\#\#\#END2\#\#\#" as last line.
     \\

\bottomrule
\end{tabular}}
\end{table*}

\clearpage
\begin{table*}

\caption{\textbf{Corrective  re-prompting template} used in the \method \emph{Correct}  module. The prompt is structured into fixed and dynamic components. The fixed components remain constant across all samples, while dynamic fields are populated based on the violation instance. (For semantic violations involving visual content, an additional instruction guides the LLM to reason over Webpage screenshots—see purple-highlighted rows.) }
\label{fig:mp_rePrompt}
\renewcommand{\arraystretch}{1.5}
\rowcolors{2}{white}{lightgray}
\centering
\resizebox{\textwidth}{!}{
\begin{tabular}{p{2.3cm}p{1cm}p{14cm}}
\toprule
\textbf{Metacognitive Prompting Stage} & \textbf{Type} & \textbf{Prompt Template} \\
\midrule

\multirow{2}{*}{\makecell{Comprehension \\  Clarification} }
& Fixed & \begin{minipage}[t]{\linewidth}
\ttfamily\small
  \texttt{\{Role-play persona\}} + You are analyzing a Web accessibility issue using a snippet of Affected HTML Element(s) \colorbox{blue!10}{, Web page screenshot} and related metadata. \colorbox{blue!10}{The screenshot reflects exactly what is rendered to users.}
  Follow these strict rules:\\
  \begin{itemize}
  \item \colorbox{blue!10}{Prioritize visual analysis: list at least three specific details observable in the image} \colorbox{blue!10}{(e.g., color, shape, text, or spatial arrangement).}
    \item Analyze only the provided HTML snippet and metadata. Do not infer or invent additional structure, styles, or UI elements beyond what is given.
    
    \item Avoid introducing or rewriting content not present in the HTML. Do not add or alter CSS, forms, headers, sections, scripts, or attributes unnecessarily. Modify only the minimal code needed to resolve the violation.
    
    \item Return only the modified lines in a fenced code block. Leave all other parts of the HTML unchanged.
    
    \item Justify every accessibility concern directly with observable evidence from the HTML.

\end{itemize}
Clarify your understanding of the following Web accessibility violation: 

\end{minipage} \textcolor{lightgray}{Lorem Ipsum is simply dummy text of the printing and typesetting industry.}\\
& Dynamic & \texttt{\{Category\}} \newline
\texttt{\{Category description\}} \newline
     \texttt{\{Violation name\}} \newline
      \texttt{\{Violation description\}} \newline
     \texttt{\{URL\}} \newline
     \texttt{\{HTMLElement\}}      \newline 
     \texttt{\{Impact\}}  \newline
     \colorbox{blue!10}{\texttt{\{Web page screenshot\}}}
     \\
     & Fixed & Impact is a rating determined by the severity of the violation, indicating the extent to which it hinders user interaction with the Web content. The scale is [cosmetic, minor, moderate, serious, critical]\\
     \hline

\multirow{2}{*}{\makecell{Preliminary \\  Judgment}} 
& Fixed &   Based on your understanding, provide a preliminary correction for the Web accessibility violation based on the following WCAG guideline(s): \textcolor{white}{violation based on the following WCAG guideline(s):}\\ 
& Dynamic & \texttt{\{Relevant WCAG guideline\}} \\ 
&Fixed &  Make sure your generated code corrects the Web accessibility violation without introducing new accessibility violations. Ensure you generate the complete corrected code, not just a snippet. \\ \hline

\multirow{2}{*}{\makecell{Critical \\ Evaluation} }
& Fixed &     Critically assess your preliminary correction, make sure to correct the initial Web accessibility violation without introducing new Web accessibility violations. Only make corrections if the previous answer is incorrect. Make sure your generated code corrects the Web accessibility violation without introducing new accessibility violations.\\ \hline

\multirow{2}{*}{\makecell{Decision \\ Confirmation}} 
& Fixed &  Confirm your final decision on whether the correction is accurate or not, and provide the reasoning for your decision. Only suggest further corrections if the initial response contains errors. Make sure your generated code corrects the Web accessibility violation without introducing new accessibility violations. Enclose your corrected HTML code to replace the initial code with accessibility violations between these two marker strings: "\#\#\#START\#\#\#" as the first line and "\#\#\#END\#\#\#" as the last line.
  \\ \hline

\multirow{2}{*}{\makecell{Confidence \\ Assessment} }
& Fixed & Evaluate your confidence (0-100\%) in your correction, enclose your confidence score between these two marker strings: "\#\#\#START1\#\#\#" as first line and "\#\#\#END1\#\#\#" as last line. Provide an explanation for this confidence level; enclose your explanation between these two marker strings: "\#\#\#START2\#\#\#" as the first line and "\#\#\#END2\#\#\#" as last line.
     \\

\bottomrule
\end{tabular}}
\end{table*}

\clearpage
\renewcommand{\arraystretch}{1.3} 
\begin{table*}
\caption{Prompt templates used for \textbf{baseline methods}, structured into fixed and dynamic components. Fixed components remain constant across all samples, while dynamic fields are populated based on the specific violation instance. These templates are directly adapted from the original works. (For semantic violations involving visual content, we added an additional instruction that guides the LLM to reason over Webpage screenshots—see purple-highlighted rows.)}
\label{tab:baseline_prompt_templates}
\resizebox{\textwidth}{!}{
\rowcolors{2}{white}{lightgray}
\begin{tabular}{p{2.5cm} p{1cm} p{13cm}}
\toprule
\textbf{Baseline Method} & \textbf{Type} & \textbf{Prompt Template} \\
\midrule

\multirow{2}{*}{\parbox[t]{3.2cm}{\textbf{Contextual} \\ \textbf{Prompting} \\ (Othman et al.\cite{DBLP:conf/petra/OthmanDJ23})}} 
& Fixed & Given the following \colorbox{blue!10}{Web page screenshot and} source code, can you fix the accessibility issue related to the success criteria according to WCAG 2.1? \textcolor{lightgray}{Given the following source code, can you fix the accessibility issue related to the success criteria according to WCAG 2.1? } \\
& Dynamic & \texttt{\{HTML\}}, \texttt{\{WCAG relevant to the violation\}}  \colorbox{blue!10}{, \texttt{\{Web page screenshot\}}} \\ 
\hline

\multirow{2}{*}{\parbox[t]{3.2cm}{\textbf{ReAct Prompting} \\ (Huang et al.~\cite{DBLP:journals/corr/abs-2401-16450})}} 
& Fixed & 
\begin{tabular}[t]{@{}l@{}}
You are a helpful assistant who will correct accessibility issues of a provided Website. \\
Provide your thought before 
 you provide a fixed version of the results. \\
 E.g. \textbf{Incorrect:} <span>Search</span> \\
\textbf{Thought:} because ... I will ... \\
\textbf{Correct:} <span class="DocSearch-Button-Placeholder">Search</span> \\
You are operating on this Website: 
\end{tabular} \\
& Dynamic & \makecell{\texttt{\{Web page URL\}},  \texttt{\{violation name\}}, 
\texttt{\{violation description\}}, 
\texttt{\{fix advice\}}, 
\texttt{\{HTML\}}}
\\
\rowcolor{blue!10} &Fixed & Given the Web page screenshot: \\
\rowcolor{blue!10} &Dynamic & \texttt{\{Web page screenshot\}} \\
\hline

\multirow{2}{*}{\parbox[t]{3.2cm}{\textbf{Zero-shot} \\ \textbf{Prompting} \\(Delnevo et al.~\cite{DBLP:conf/ccnc/DelnevoAM24})}} 
& Fixed & Is the following HTML code accessible? \textcolor{lightgray}{Is the following HTML code accessible?}
\textcolor{lightgray}{Is the following HTML code accessible?}\textcolor{lightgray}{Is the following HTML code accessible?}\textcolor{lightgray}{Is the following HTML code accessible?}\textcolor{lightgray}{Is the following HTML code accessible?}\\

& Dynamic & \texttt{\{HTML\}}  \\
\rowcolor{blue!10} &Fixed & Given the Web page screenshot: \\
\rowcolor{blue!10} &Dynamic & \texttt{\{Web page screenshot\}} \\
\bottomrule
\end{tabular}}
\end{table*}

\clearpage

\begin{table*}
\centering
\caption{Subset of our Proposed Taxonomy to Categorize Web Accessibility Violations. The full taxonomy can be found in our supplementary material.}
\resizebox{\textwidth}{!}{
\begin{tabular}{|p{1.8cm}|p{2cm}|p{7cm}|p{1.5cm}|p{1.5cm}|p{2.7cm}|}
\hline
\textbf{Category} & \textbf{Violation Name} & \textbf{Description} & \textbf{Violated Guidelines} & \textbf{Impact} & \textbf{Supplementary Information} \\ \hline

\multirow{5}{*}{\shortstack{\textbf{Semantic} \\ \textbf{Accessibility} \\\textbf{Violations}}}

& image-alt-not-descriptive      & Inaccurate or misleading alternative text that fails to describe the purpose of the image.            & WCAG 1.1.1             &  Critical & Image   \\ \cline{2-6}
& video-captions-not-descriptive    & Inaccurate video captions.                                 & WCAG 1.2.1, 1.2.3      & Critical  &  Video \\ \cline{2-6}
& lang-mismatch      & Page language attribute does not match the actual language of the content.                            & WCAG 3.1.1             & Serious  &   -- \\ \cline{2-6}

      & link-text-mismatch  & Links fail to convey their purpose or are ambiguous.                                                  & WCAG 2.4.4, 2.4.9      & Serious  &   -- \\ \cline{2-6}
      & button-label-mismatch       & Button labels are unclear or fail to specify their purpose.                                            & WCAG 4.1.2, 2.5.3             &  Critical  &  -- \\ \cline{2-6}
      & ambiguous-heading  & Headings are vague, repetitive, or fail to describe the content.                                      & WCAG 2.4.6, 2.4.10     &  Moderate  &   -- \\ \cline{2-6}
      & incorrect-semantic-tag    & A non-semantic tag (e.g., div or span) is used instead of a proper semantic element (e.g., header, nav, main).                                      & WCAG 1.3.1             & Serious  &  Document structure (other headings, section context) \\ \cline{2-6}
     
       & color-only-distinction            & Visual information is conveyed using color alone without additional indicators like text, shape, or pattern, making it inaccessible to users with color vision deficiencies. & WCAG 1.4.1             &  Serious    & Web page Screenshot   \\ \hline

\multirow{5}{*}{\shortstack{\textbf{Layout} \\ \textbf{Accessibility}\\ \textbf{ Violations}}}

&meta-viewport      & Ensure <meta name="viewport"> does not disable text scaling and zooming            &    WCAG 1.4.4         &  Critical & -- \\ \cline{2-6}

&color-contrast      &    Ensure the contrast between foreground and background colors meets WCAG 2 AA minimum contrast ratio thresholds         &       WCAG 1.4.3     &  Serious & Color Information (Background and Foreground)\\ \cline{2-6}

&avoid-inline-spacing      &    Ensure that text spacing set through style attributes can be adjusted with custom stylesheets         &     WCAG 1.4.12        &  Serious & --\\ \cline{2-6}
&target-size      &     Ensure touch targets have sufficient size and space        &        WCAG 2.5.5     & Serious  & --\\ \hline

\multirow{5}{*}{\shortstack{\textbf{Syntax} \\ \textbf{Accessibility} \\ \textbf{ Violations}}}

& duplicate-id-aria      &       Ensure every id attribute value used in ARIA and in labels is unique      &    WCAG 4.1.2      &  Critical   &  -- \\ \cline{2-6}

& tabindex      &       Ensure tabindex attribute values are not greater than 0      &         WCAG 2.1.1     &  Serious  &  -- \\ \cline{2-6}

& duplicate-id-aria      &       Ensure every id attribute value used in ARIA and in labels is unique      &       WCAG 4.1.2      &  Critical   &  -- \\ \cline{2-6}

& tabindex      &       Ensure tabindex attribute values are not greater than 0      &          WCAG 2.1.1     &  Serious  &  -- \\ \cline{2-6}

& valid-lang      &       Ensure lang attributes have valid values      &    3.1.2       &  Serious  &  -- \\ \cline{2-6}

& aria-required-attr      &      Ensure elements with ARIA roles have all required ARIA attributes       &      WCAG 4.1.2       &  Critical   &  -- \\ \cline{2-6}

& meta-refresh      &       Ensure <meta http-equiv="refresh"> is not used for delayed refresh      &      WCAG 2.2.1       & Critical &  -- \\ \cline{2-6}

& empty-table-header      & Ensure table headers have discernible text            &   WCAG 1.3.1, 2.4.6          &  Minor & -- \\ \cline{2-6}

& empty-heading      &  Ensure headings have discernible text           &      WCAG 1.3.1, 2.4.6       & Minor  & -- \\ \hline

\end{tabular}
\label{tab:important_violations}}
\end{table*}

\end{document}